\documentclass[reprint,superscriptaddress,nofootinbib,floatfix,float,aps, prd]{revtex4-1}
\usepackage{amsmath}
\usepackage{amssymb}
\usepackage{graphicx}
\usepackage{dcolumn}
\usepackage{bm}
\usepackage[colorlinks=true, allcolors=blue]{hyperref}
\usepackage{cleveref}
\usepackage{xcolor}
\usepackage[normalem]{ulem}
\begin{document}

\title[]{Gravitational Waves Emission in Quadratic Gravity: longitudinal
modes, angular momentum emission, and positivity of the radiated power}

\author{Matheus F. S. Alves}
 \email{matheus.s.alves@edu.ufes.br}
\affiliation{Departamento de Física, Universidade Federal do Espírito Santo, Vitória, ES,  29075-910, Brazil.}
\affiliation{Núcleo de Astrofísica e Cosmologia (Cosmo-Ufes), Universidade Federal do Espírito Santo, Vitória, ES,  29075-910, Brazil.}

\author{R.R. Cuzinatto}
\email{rodrigo.cuzinatto@unifal-mg.edu.br}
\affiliation{Instituto de Ci\^encia e Tecnologia, Universidade Federal de
Alfenas, Rodovia Jos\'e Aur\'elio Vilela, 11999, Cidade Universit\'aria,
CEP 37715-400 Po\c cos de Caldas, Minas Gerais, Brazil.}

\author{C.A.M. de Melo}
\email{cassius.melo@unifal-mg.edu.br}
\affiliation{Instituto de Ci\^encia e Tecnologia, Universidade Federal de
Alfenas, Rodovia Jos\'e Aur\'elio Vilela, 11999, Cidade Universit\'aria,
CEP 37715-400 Po\c cos de Caldas, Minas Gerais, Brazil.}

\author{L.G. Medeiros}
\email{leo.medeiros@ufrn.br}
\affiliation{Escola de Ci\^encia e Tecnologia, Universidade Federal do Rio Grande
do Norte, Campus Universit\'ario, s/n\textendash Lagoa Nova, CEP 59078-970,
Natal, Rio Grande do Norte, Brazil.}

\author{P.J. Pompeia}
\email{pompeia@ita.br}
\affiliation{Departamento de F\'isica, Instituto Tecnol\'ogico da Aeron\'autica,
Pra\c ca Mal. Eduardo Gomes, 50, CEP 12228-900 S\~ao Jos\'e dos Campos, S\~ao
Paulo, Brazil.}

\begin{abstract}
In this paper, the emission of gravitational waves in quadratic gravity theory is examined. The wave equations for massless and massive perturbations are derived, followed by the calculation of the energy and angular momentum radiated. In the quadrupole approximation, and taking into account only the transverse-traceless modes, it is shown that the theory avoids the issues generated by the Ostrogradsky instabilities and achieves positive energy and angular momentum emissions. As an example, a rotating ellipsoid with free precession is analyzed, and the effects of the massive perturbations on its rotation are highlighted.
\end{abstract}

\maketitle

\section{Introduction\label{sec:Introduction}}

Gravitational waves are certainly one of the most disruptive discoveries in
astrophysics in recent years. Although its prediction dates back to 1916
\cite{EinsteinGW1916}, the first evidence of its existence was only
glimpsed in the 1970s by Hulse and Taylor \cite{HulseTylor1975}, who
observed that the system PSR 1913+16 lost energy and angular momentum in
accordance with the theory of General Relativity (GR). It took another 40
years for the first direct detection of a gravitational wave (GW), due to
the merger of two black holes, by the LIGO collaboration in 2015 \cite%
{Abbot2016}. This event marked the opening of a new window of observations,
creating a new perspective for understanding astrophysical phenomena. Two
years later, a major breakthrough was achieved with the observation of the
merger of two neutron stars by the LIGO-VIRGO collaboration \cite{Abbot1917}%
. This was the first event in which both the GW and its electromagnetic
counterpart were observed together, marking the beginning of the era of
multimessenger astronomy.

Besides allowing the comprehension of astrophysics phenomena from a broader
perspective, GW is also a powerful tool for studying the gravitational
interaction itself. Although GR is the standard theory for gravitation, with
several successful predictions, it fails to describe galaxy rotation
curves and the accelerated expansion of the Universe in the presence of
ordinary matter \cite{Hooper2018,Caldwell2009,Astier2006,Riess2004}; the
consistent description of the early (inflationary) Universe is hardly
achieved by GR without the presence of extra degrees of freedom \cite%
{Starobinsky1980,Guth1981,Enciclopedia2014,NosJCAP2019,SilSobMed2022,BezerraLeo2022,GesielLeo2022}%
. Also, the non-renormalizability and the presence of singularities \cite%
{UtiyamaDeWitt1962,Stelle1977} do not allow us to have a self-consistent
quantum version of GR with the standard (perturbative) prescription of
quantization. Due to these limitations, there is almost a consensus on the
scientific community that GR should not be taken as the definitive theory of
gravitation. This motivated the search for new theories of gravity.

The proposition of new theories of gravitation can be done by a plethora of
different approaches. They include the modification of the geometry \cite%
{HayashiShirafuji1979,Pereira2004,Cassius2007,Cuzinatto2021, Cuzinatto:2006hb,Battista2021,Battista2022}, the
modification of the gravitational action \cite%
{Faraoni2010,Capozziello2011,Novello2013,Max2014a,Romero2016a,Romero2016b,
NosPRD2016,NosArxiv2018,NOJIRI201159,NOJIRI20171,Ghosh2021}
(in the Riemann manifold), and many others. Usually, all these proposals
share one same concern: one should obtain GR (or GR-equivalent theories) as
a limiting case. With this in mind, we can understand that one of the
simplest extensions to GR, besides considering the cosmological constant,
consists of considering contributions to the gravitational action integral
of terms that are quadratic in the curvature (this is, for instance, the
case of the Starobinsky model of inflation \cite{Starobinsky1980}). The
presence of quadratic terms in the action (quadratic gravity), however,
brings with them a significant modification of the dynamical equations: the
field equations are of fourth order in the derivative of the metric tensor.

Theories with higher-order derivatives are quite appealing in the context of
quantum gravity, and some of the problems of GR may eventually be worked
around -- for instance, it was shown by Stelle \cite{Stelle1977}, that a
renormalizable theory of gravity can be obtained for the quadratic
Lagrangian. Despite being renormalizable, quadratic gravity exhibits
Ostrogradsky instability, which, when quantized, gives rise to ghost-like
fields \cite{Sbisa2015}. The quantization of these fields is a nontrivial
task and often leads to issues with unitarity \cite{PaisUhl1950}. One way to
address this issue is to recognize that ghosts are unstable and, as a
result, do not contribute to the asymptotic spectrum of the theory \cite%
{Modesto2016,Donoghue2019}. Alternatively, one could define a nontrivial
norm for ghost states to restore their unitarity \cite{Salvio2019,Salvio2021}%
. While there is currently no consensus in the literature on how to handle
the ghosts present in higher-order models, quadratic gravity and its
extensions \cite{RacLeoPinSha2021,Asorey1997,Modesto2016,Rachwal2022} can be
treated as effective theories that implement corrections to GR.

The investigation of gravitational waves within the framework of quadratic
gravity was particularly analyzed in Ref. \cite{Alves2022}.\footnote{%
See also Refs. \cite%
{Ody2003,Bogdanos2010,Capoz2017,Holscher2019,Kim2019,Avijit2022,Odin2022,Marcio2024} for
related works.} In this study, a modification to the Einstein-Hilbert action
was proposed, incorporating quadratic contributions from the scalar
curvature and the Weyl tensor, i.e., $R^{2}$ and $C_{\mu\nu\alpha\beta}C^{\mu%
\nu \alpha\beta}$ terms, respectively. The GWs obtained in this work
exhibited two spin-2 modes---one massive and one massless---along with a
scalar (spin-0) mode. The emission of GWs from an inspiraling binary system
composed of two black holes in the non-relativistic and circular orbit
approximations was analyzed, and the parameters of the model were
constrained by observational data. Although this work yielded interesting
results, it neglected the contributions of the massive spin-2 longitudinal
polarizations. Additionally, it did not investigate the loss of angular
momentum due to GW emission in the context of quadratic gravity. The present
paper aims to address these two issues.

The non-conservation of angular momentum in gravitational systems plays an
important role both in the dynamics of the physical system that generates
the GWs and in the dynamics of the GWs themselves. This is a key focus of
the present paper: evaluating the effect of angular momentum transfer from
the source to the GWs. This analysis requires careful consideration, as in
the linear regime, quadratic actions behave like classical field theories
with Lagrangians that present second-order derivatives. From this
perspective, we cannot apply the known formulas for the angular momentum
current usually derived in the first-order case. 
Specifically, the relation between the energy ($E$) loss and angular momentum ($J$) loss is given by $\dot{E}=\omega\dot{J}$ for quasi-circular orbits ($\omega$ is the angular velocity and an overdot denotes a time derivative). We will see below in this paper that this relation is not exactly valid in the context of quadratic gravity. This will demand a proper extension, which will be presented here. 

In the context of GWs, the term $C_{\mu\nu\alpha\beta}C^{\mu\nu\alpha\beta}$%
, associated with the massive spin-$2$ mode, encompasses five degrees of
freedom: two transverse and three longitudinal (two of which are linearly dependent with respect to the remaining degrees of freedom of the metric theory due to covariance requirements--more on this below in the text). Furthermore,
this term is also responsible for introducing the Ostrogradsky instability,
which is evident through a negative sign in the energy density carried away
by the gravitational wave. Focusing on the specific case of binary systems
in the circular orbit approximation, Ref. \cite{Alves2022} demonstrated that
it is possible to consistently manage this instability by restricting the
analysis to the transverse degrees of freedom. However, does this
consistency extend to more general scenarios that include longitudinal
polarizations and/or account for angular momentum loss? Addressing this
question is the main motivation for this paper.

In the present paper, we intend to generalize the analysis of the emission of
GWs in quadratic gravity presented in Ref. \cite{Alves2022}. We will start
by presenting, in Section II, the linear approximation for our quadratic
model and the resulting wave equations for the perturbations. Therein, we
obtain the conserved currents for our system which are related to the
energy-momentum tensor and angular momentum current, both for the GWs. Next,
we evaluate the energy and angular momentum carried away by the GWs. In
Section III, we restrict the analysis to the quadrupole approximation and
consider the propagation of GWs in a generic direction. Then, we evaluate
the power radiated and consider a first example of a binary system in
circular orbit. This section also includes a detailed discussion of the
Ostrogradsky instability and a possible solution to this issue. In Section
IV, the source of GWs is taken as a rotating ellipsoid with free precession,
and the role played by the quadratic terms is analyzed in comparison with
the predictions of GR. At last, in Section V, we present our final comments.

\section{Gravitational waves in quadratic gravity\label{sec:GWQG}}

The action integral of quadratic gravity (QG) is given by
\begin{eqnarray}
S=-\frac{1}{2\kappa}\int d^{4}x\sqrt{-g}\left[R+\frac{1}{2}\gamma R^{2}-%
\frac{1}{2}\alpha C_{\mu\nu\alpha\beta}C^{\mu\nu\alpha\beta}\right] \nonumber \\
+\int
d^{4}x\sqrt{-g}\mathcal{L}_{m},  \ \ \ \ \ \ \  \ \ \ \ \ \ \  \ \ \ \ \ \ \   \ \ \ \ \ \ \  \ \ \ \ \ \ \  \ \ \ \ \ \ \ \label{eq:S}
\end{eqnarray}
where $C_{\mu\nu\alpha\beta}$ is the Weyl tensor, $\mathcal{L}_{m}$ is the
matter Lagrangian density, $\kappa=8\pi G/c^{4}$ and $\alpha$, $\gamma$ are
parameters with dimension $\left(\text{length}\right)^{2}$.

In this paper, we shall carefully study the process of emission and
propagation of gravitational waves in a flat background in the context of
QG. We write the metric tensor, as usual, \cite{Maggiore2007},
\begin{equation}
g_{\mu\nu}=\eta_{\mu\nu}+\left(\bar{h}_{\mu\nu}-\frac{1}{2}\eta_{\mu\nu}\bar{%
h}\right),  \label{eq:g(h-bar)}
\end{equation}
and, inspired by Ref. \cite{Alves2022}, decompose $\bar{h}_{\mu\nu}$ in the
form
\begin{equation}
\bar{h}_{\mu\nu}=\tilde{h}_{\mu\nu}+\Psi_{\mu\nu}-\eta_{\mu\nu}\left(\Phi+%
\frac{1}{6}\Psi\right).  \label{eq:h-bar(h-tilde,Psi,Phi)}
\end{equation}
Here $\bar{h}=\eta^{\mu\nu}\bar{h}_{\mu\nu}$ and $\Psi=\eta^{\mu\nu}\Psi_{%
\mu\nu}$. By the end of this section, it should be clear that the
decomposition in Eq. (\ref{eq:h-bar(h-tilde,Psi,Phi)}) allows to recognize $%
\tilde{h}_{\mu\nu}$ as the regular massless spin-$2$ field, $\Phi$ as a
massive spin-$0$ field, and $\Psi_{\mu\nu}$ as a massive spin-$2$ field. The
fields $\Phi$ and $\Psi_{\mu\nu}$ are the additional contribution of QG for
GW physics.

The propagation of GW is studied after one approximates the invariants in
Eq. (\ref{eq:S}) up to second order in $\bar{h}_{\mu\nu}$ and its
derivatives. After a long but straightforward calculation, we obtain
\begin{align}
\sqrt{-g}R & = -\frac{1}{4}\partial _{\mu }\tilde{h}_{\alpha \beta }\partial
^{\mu }\tilde{h}^{\alpha \beta }-\frac{1}{4}\partial _{\mu }\Psi _{\alpha
\beta }\partial ^{\mu }\Psi ^{\alpha \beta } + \frac{3}{2}\partial _{\alpha }\Phi \partial ^{\alpha }\Phi \notag \\
&  -\frac{1}{2} \partial _{\mu }\tilde{h}_{\alpha \beta }\partial ^{\mu }\Psi ^{\alpha \beta
} -\partial _{\beta }\tilde{h}^{\alpha \beta }\partial _{\alpha }\Phi -\partial _{\beta }\Psi ^{\alpha \beta }\partial _{\alpha }\Phi \notag \\
& -\frac{1}{2}\partial _{\mu }\tilde{h}\partial ^{\mu }\Phi +\frac{1}{2}%
\partial _{\alpha }\Phi \partial ^{\alpha }\Psi ,  \label{eq:R}
\end{align}
\begin{align}
\sqrt{-g}R^{2} & = 9\left( \square \Phi \right) ^{2}-6\partial ^{\mu}\partial ^{\nu }\tilde{h}_{\mu \nu }\square \Phi   -6\partial ^{\mu }\partial ^{\nu }\Psi _{\mu \nu }\square \Phi   \notag \\
&-3\square \tilde{h}\square \Phi +3\square \Psi \square \Phi , \label{eq:R2} 
\end{align}%
and 
\begin{equation}
\sqrt{-g}C^{2} =\frac{1}{2}\square \tilde{h}_{\mu \beta }\square \tilde{h}%
^{\mu \beta }+\square \tilde{h}_{\mu \beta }\square \Psi ^{\mu \beta }
+\frac{1}{2}\square \Psi _{\mu \beta }\square \Psi ^{\mu \beta },  \label{eq:C2}
\end{equation}
where $\square=\eta^{\mu\nu}\partial_{\mu}\partial_{\nu}$ and $
C^{2}=C_{\mu \nu \alpha \beta }C^{\mu \nu \alpha \beta }$. Eqs. (\ref{eq:R}%
), (\ref{eq:R2}), and (\ref{eq:C2}) assume the
harmonic-transverse-traceless-like gauge \cite{Alves2022}
\begin{equation}
\partial^{\mu}\tilde{h}_{\mu\nu}=\partial^{\mu}\Psi_{\mu\nu}=0\qquad\text{and%
}\qquad\tilde{h}=\Psi=0.  \label{eq:harmonic-TT-gauge}
\end{equation}
It is paramount that the gauge fixing occurs after the functional
derivatives are taken. The terms involving the quantities in (\ref%
{eq:harmonic-TT-gauge}) are, therefore, neglected henceforth.

Accordingly, the QG action (\ref{eq:S}) in vacuum $\left(\mathcal{L}%
_{m}=0\right)$ reduces to:
\begin{equation}
\bar{S}=-\frac{1}{2\kappa}\int d^{4}x\mathcal{L},  \label{eq:2ndOrderS}
\end{equation}
with
\begin{align}
\mathcal{L}& =\frac{1}{2\kappa }\left[ \frac{1}{4}\partial _{\mu }\tilde{h}%
_{\alpha \beta }\partial ^{\mu }\tilde{h}^{\alpha \beta }+\frac{1}{4}%
\partial _{\mu }\Psi _{\alpha \beta }\partial ^{\mu }\Psi ^{\alpha \beta}  -\frac{3}{2}\partial _{\alpha }\Phi \partial ^{\alpha }\Phi \right.  \notag \\
& \qquad +\frac{1}{2}\partial _{\mu }\tilde{h}_{\alpha \beta }\partial ^{\mu }\Psi ^{\alpha \beta} +\partial _{\beta }\tilde{h}^{\alpha \beta }\partial _{\alpha }\Phi
+\partial _{\beta }\Psi ^{\alpha \beta }\partial _{\alpha }\Phi   \notag \\
& \qquad \left. +\frac{1}{2}\partial _{\mu }\tilde{h}\partial ^{\mu }\Phi -%
\frac{1}{2}\partial _{\alpha }\Phi \partial ^{\alpha }\Psi \right]   \notag
\\
& +\frac{\gamma }{2\kappa }\left[ -\frac{9}{2}\left( \square \Phi \right)
^{2}+3\partial _{\alpha }\partial _{\beta }\tilde{h}^{\alpha \beta }\square
\Phi +3\partial _{\alpha }\partial _{\beta }\Psi ^{\alpha \beta }\square \Phi \right.   \notag \\
& \qquad \left. + \frac{3}{2}\square \tilde{h}\square \Phi   -\frac{3}{2}\square \Psi \square \Phi \right]   \notag \\
& +\frac{\alpha }{2\kappa }\left[ \frac{1}{4}\square \tilde{h}_{\alpha \beta
}\square \tilde{h}^{\alpha \beta }+\frac{1}{2}\square \tilde{h}_{\alpha
\beta }\square \Psi ^{\alpha \beta } +\frac{1}{4}\square \Psi _{\mu \beta }\square \Psi ^{\mu \beta }%
\right]. \label{eq:2ndOrderL} 
\end{align}
The next step is to calculate the vacuum field equations. By extremizing $%
\bar{S}$ in (\ref{eq:2ndOrderS}) with respect to the fields $\Phi$, $\tilde{h%
}_{\mu\nu}$ and $\Psi_{\mu\nu}$ while making use of the gauge conditions (%
\ref{eq:harmonic-TT-gauge}), we obtain, after some degree of manipulation:
\begin{align}
3\gamma\square\Phi-\Phi & =0,  \label{eq:EOM-Phi} \\
\square\left(\tilde{h}_{\alpha\beta}+\Psi_{\alpha\beta}-\alpha\square\tilde{h%
}_{\alpha\beta}-\alpha\square\Psi_{\alpha\beta}\right) & =0.
\label{eq:EOM-h-tilde-Psi}
\end{align}
Eq. (\ref{eq:EOM-Phi}) is obviously of the Klein-Gordon type. The tensorial
equation (\ref{eq:EOM-h-tilde-Psi}) admits a massive mode and a massless
solution. These facts motivate the definitions:
\begin{equation}
m_{\Phi}^{2}=\frac{1}{3\gamma}\qquad\text{and}\qquad m_{\Psi}^{2}=\frac{1}{%
\alpha}.  \label{eq:mPhi_mPsi}
\end{equation}
With that, the final form of the field equation are:\footnote{
The field equations in (\ref{eq:FE-h-tilde})--(\ref{eq:FE-Psi}) are
identical to those in Ref. \cite{Alves2022} after the mapping $\alpha\rightarrow\alpha/2$. This numerical difference in the coupling factor appeared due to a small error in Ref. \cite{Alves2022}. In equation (2) of this reference, we should have $2\alpha$ instead of $\alpha$. This correction was made in the current paper and led to the new definition of $m_{\Psi}^{2}$ without the factor $2$ that appears in Ref. \cite{Alves2022}. This same correction leads to a change from $1/3$ of Eq. (2) in Ref. \cite{Alves2022} to $1/6$ in Eq. (3) of the current paper.}
\begin{align}
\square\tilde{h}_{\alpha\beta} & =0,  \label{eq:FE-h-tilde} \\
\left(\square-m_{\Phi}^{2}\right)\Phi & =0,  \label{eq:FE-Phi} \\
\left(\square-m_{\Psi}^{2}\right)\Psi_{\alpha\beta} & =0.  \label{eq:FE-Psi}
\end{align}
The structure of the above equations characterizes $\Phi$, $\tilde{h}%
_{\alpha\beta}$ and $\Psi_{\alpha\beta}$ as a massive scalar field, a
massless tensorial field, and a massive tensorial field, respectively.
Notice that the massive degrees of freedom vanish in the limit $%
\alpha,\gamma\rightarrow0$; in which case the familiar results from general
relativity are recovered.

\subsection{Conserved currents\label{subsec:Conserved-currents}}

The previous construction leading to the Lagrangian density (\ref%
{eq:2ndOrderL}) allows us to interpret quadratic gravity as a classical
field theory. In this context, the quantity $\bar{h}_{\mu\nu}$ is understood
as a classical field in a Minkowski background.

Accordingly, infinitesimal transformations take on the form \cite%
{Maggiore2007}
\begin{align}
x^{\prime\mu} & =x^{\mu}+\epsilon^{a}A_{a}^{\mu}\left(x\right),
\label{eq:coord-transf} \\
\bar{h}_{\alpha\beta}^{\prime}\left(x^{\prime}\right) & =\bar{h}%
_{\alpha\beta}\left(x\right)+\epsilon^{a}F_{\alpha\beta,a}\left(\bar{h}%
_{\alpha\beta},\partial\bar{h}_{\alpha\beta},\partial^{2}\bar{h}%
_{\alpha\beta}\right),  \label{eq:field-transf}
\end{align}
where $\epsilon^{a}$ with $a=1,...,N$ are the transformation parameters. The
functions $A_{a}^{\mu}$ and $F_{\alpha\beta,a}$ specify the transformation
rules for the coordinates $x^{\mu}$ and the field $\bar{h}_{\alpha\beta}$,
respectively.

The field theory formalism teaches us that symmetry transformations leaving
the action invariant produce conserved quantities associated with Noether
currents. These currents can be determined from the specific Lagrangian
density at hand and from the functions $A_{a}^{\mu}$ e $F_{\alpha\beta,a}$
as follows.
Usually, one deals with first-order Lagrangian densities, i.e. those depending on the 
the field and its first-order derivatives. However, quadratic gravity requires a second-order Lagrangian, viz.
$\mathcal{L}=
\mathcal{L}\left(\bar{h}_{\alpha\beta},\partial_{\mu}\bar{h}_{\alpha\beta},
\partial_{\mu}\partial_{\nu}\bar{h}_{\alpha\beta}\right)
$. As a consequence, the structural form of the current in terms of $\mathcal{L}$ must be generalized. The expression for the conserved current $j_{a}^{\mu}$ related to the second-order Lagrangian is derived in Appendix \ref{app:Current}; it reads:
\begin{align}
j_{a}^{\mu } & = \frac{\partial \mathcal{L}}{\partial \partial _{\mu }\bar{h}
_{\alpha \beta }}\left( A_{a}^{\sigma }\partial _{\sigma }\bar{h}_{\alpha
\beta }-F_{\alpha \beta ,a}\right) -A_{a}^{\mu }\mathcal{L} \notag \\
& -2\partial _{\nu }\frac{\partial \mathcal{L}}{\partial \partial _{\mu
}\partial _{\nu }\bar{h}_{\alpha \beta }}\left( A_{a}^{\sigma }\partial _{\sigma }
\bar{h}_{\alpha \beta }-F_{\alpha \beta ,a}\right) \notag \\
& +\partial _{\nu }\left\{ \frac{\partial \mathcal{L}}{\partial \left(
\partial _{\mu }\partial _{\nu }\bar{h}_{\alpha \beta }\right) }\left(
A_{a}^{\sigma }\partial _{\sigma }\bar{h}_{\alpha \beta }-F_{\alpha \beta
,a}\right) \right\} . \label{eq:conserved-current} 
\end{align}
The Noether currents associated to gravitational waves are meaningful when
computed as spacetime averages $\left\langle j_{a}^{\mu}\right\rangle $. For
this reason, boundary terms are sub-dominant and are neglected in a
first-order approximation. Accordingly, the last term of Eq. (\ref%
{eq:conserved-current}) will be disregarded throughout this work.\footnote{For economy in notation, the average symbol $\left\langle ...\right\rangle $
will be written explicitly in the final results only.}
The decomposition of the metric $\bar{h}_{\alpha\beta}$ in terms of the
fields $\tilde{h}_{\alpha\beta}$, $\Psi_{\alpha\beta}$ and $\Phi$, cf. Eq. (%
\ref{eq:h-bar(h-tilde,Psi,Phi)}), enables us to compute the current as the
sum of partial contributions from each of the latter field:\footnote{The linear combination in Eq. (\ref{eq:total-current}) is actually possible
because each one of the fields $\tilde{h}_{\alpha\beta}$, $%
\Psi_{\alpha\beta} $ and $\Phi$ satisfies an independent equation of motion.}
\begin{equation}
j_{a}^{\mu}=\left(j_{a}^{\mu}\right)^{\tilde{h}}+\left(j_{a}^{\mu}\right)^{%
\Psi}+\left(j_{a}^{\mu}\right)^{\Phi} -A_{a}^{\mu}  \mathcal{L}
\label{eq:total-current}
\end{equation}
with
\begin{align}
\left( j_{a}^{\mu }\right) ^{\tilde{h}}& =\frac{\partial \mathcal{L}}{\partial \partial _{\mu }\tilde{h}_{\alpha \beta }}\left( A_{a}^{\sigma
}\partial _{\sigma }\tilde{h}_{\alpha \beta }-F_{\alpha \beta ,a}\right) \notag \\
& -2\partial _{\nu }\frac{\partial \mathcal{L}}{\partial \partial _{\mu
}\partial _{\nu }\tilde{h}_{\alpha \beta }}\left( A_{a}^{\sigma }\partial
_{\sigma }\tilde{h}_{\alpha \beta }-F_{\alpha \beta ,a}\right) ,  \label{eq:spin-2-massless-current}  \\
\left( j_{a}^{\mu }\right) ^{\Psi }& =\frac{\partial \mathcal{L}}{\partial
\partial _{\mu }\Psi _{\alpha \beta }}\left( A_{a}^{\sigma }\partial
_{\sigma }\Psi _{\alpha \beta }-F_{\alpha \beta ,a}\right) \notag \\
& -2\partial _{\nu }\frac{\partial \mathcal{L}}{\partial \partial _{\mu
}\partial _{\nu }\Psi _{\alpha \beta }}\left( A_{a}^{\sigma }\partial
_{\sigma }\Psi _{\alpha \beta }-F_{\alpha \beta ,a}\right) , \label{eq:spin-2-massive-current} \\
\left( j_{a}^{\mu }\right) ^{\Phi }& =\frac{\partial \mathcal{L}}{\partial
\partial _{\mu }\Phi }\left( A_{a}^{\sigma }\partial _{\sigma }\Phi
-F_{\alpha \beta ,a}\right)  \notag \\
& -2\partial _{\nu }\frac{\partial \mathcal{L}}{\partial \partial _{\mu
}\partial _{\nu }\Phi }\left( A_{a}^{\sigma }\partial _{\sigma }\Phi
-F_{\alpha \beta ,a}\right) .  \label{eq:spin-0-massive-current}
\end{align}

The next step is to calculate the various functional derivatives of $%
\mathcal{L}$ in Eqs. (\ref{eq:spin-2-massless-current})--(\ref%
{eq:spin-0-massive-current}) and simplify the results by utilizing the
harmonic gauge in (\ref{eq:harmonic-TT-gauge}) and the field
equations---Eqs. (\ref{eq:FE-h-tilde}), (\ref{eq:FE-Psi}), (\ref{eq:FE-Phi}%
). This task is time-consuming but straightforward; it yields:

\begin{align}
\left( j_{a}^{\mu }\right) ^{\tilde{h}}& =\frac{1}{4\kappa }\partial ^{\mu }\tilde{h}^{\alpha \beta }\left( A_{a}^{\sigma }\partial _{\sigma }\tilde{h}%
_{\alpha \beta }-F_{\alpha \beta ,a}\right) \notag \\
& -\frac{1}{4\kappa }\partial ^{\mu }\Psi ^{\alpha \beta }\left(
A_{a}^{\sigma }\partial _{\sigma }\tilde{h}_{\alpha \beta }-F_{\alpha \beta
,a}\right)   \notag \\
& -\frac{1}{4\kappa }\eta ^{\mu (\alpha }\partial ^{\beta )}\Phi \left(
A_{a}^{\sigma }\partial _{\sigma }\tilde{h}_{\alpha \beta }-F_{\alpha \beta
,a}\right)   \notag \\
& -\frac{1}{4\kappa }\eta ^{\alpha \beta }\partial ^{\mu }\Phi \left(
A_{a}^{\sigma }\partial _{\sigma }\tilde{h}_{\alpha \beta }-F_{\alpha \beta
,a}\right) ,   \label{eq:h-tilde-current}
\end{align}
\begin{align}
\left( j_{a}^{\mu }\right) ^{\Psi }& =\frac{1}{4\kappa }\partial ^{\mu }%
\tilde{h}^{\alpha \beta }\left( A_{a}^{\sigma }\partial _{\sigma }\Psi
_{\alpha \beta }-F_{\alpha \beta ,a}\right)  \notag \\
& -\frac{1}{4\kappa }\partial ^{\mu }\Psi ^{\alpha \beta }\left(
A_{a}^{\sigma }\partial _{\sigma }\Psi _{\alpha \beta }-F_{\alpha \beta
,a}\right)   \notag \\
& -\frac{1}{4\kappa }\eta ^{\mu (\alpha }\partial ^{\beta )}\Phi \left(
A_{a}^{\sigma }\partial _{\sigma }\Psi _{\alpha \beta }-F_{\alpha \beta
,a}\right)   \notag \\
& -\frac{1}{4\kappa }\eta ^{\alpha \beta }\partial ^{\mu }\Phi \left(
A_{a}^{\sigma }\partial _{\sigma }\Psi _{\alpha \beta }-F_{\alpha \beta
,a}\right) ,  \label{eq:Psi-current}
\end{align}
\begin{align}
\left( j_{a}^{\mu }\right) ^{\Phi }& =\frac{3}{2\kappa }\left[ \partial
^{\mu }\Phi \right] \left( A_{a}^{\sigma }\partial _{\sigma }\Phi -F_{\alpha
\beta ,a}\right) .  \label{eq:Phi-current}
\end{align}
In the following subsection, the conserved currents stemming from symmetries
under spacetime translations and spatial rotations will be determined.

\subsubsection{The energy-momentum tensor of the GW in quadratic gravity}

The quadratic Lagrangian (\ref{eq:2ndOrderL}) is invariant under spacetime
translations:
\begin{equation}
x^{\prime\mu}=x^{\mu}+\epsilon^{\nu}\delta_{\nu}^{\mu}\qquad\text{and}\qquad%
\bar{h}_{\alpha\beta}^{\prime}\left(x^{\prime}\right)=\bar{h}%
_{\alpha\beta}\left(x\right).  \label{eq:translations}
\end{equation}
Comparison of Eq. (\ref{eq:translations}) with Eqs. (\ref{eq:coord-transf})
and (\ref{eq:field-transf}), yields $a=\nu=0,1,2,3$,
\begin{equation}
A_{\nu}^{\mu}=\delta_{\nu}^{\mu},\qquad\text{and}\qquad
F_{\alpha\beta,\nu}=0.  \label{eq:A-F-translations}
\end{equation}

Eq. (\ref{eq:A-F-translations}) is then plugged into Eqs. (\ref%
{eq:h-tilde-current}), (\ref{eq:Psi-current}), and (\ref{eq:Phi-current}).
The resulting expressions for $\left(j_{\hphantom{\mu}\nu}^{\mu}\right)^{%
\tilde{h}}$, $\left(j_{\hphantom{\mu}\nu}^{\mu}\right)^{\Psi}$, and $%
\left(j_{\hphantom{\mu}\nu}^{\mu}\right)^{\Phi}$ are simplified recalling
the harmonic gauge condition. On top of that, we make use of fact that the
averaging processes {[}cf. comment below Eq. (\ref{eq:conserved-current}){]}
allows for integrations by parts which eliminate a few surface terms. The
results are the following:

\begin{align}
\left(j_{\hphantom{\mu}\nu}^{\mu}\right)^{\tilde{h}} & =\frac{1}{4\kappa}%
\left[\partial^{\mu}\tilde{h}^{\alpha\beta}\partial_{\nu}\tilde{h}%
_{\alpha\beta}-\partial^{\mu}\Psi^{\alpha\beta}\partial_{\nu}\tilde{h}%
_{\alpha\beta}\right],  \label{eq:j-h-tilde-transl} \\
\left(j_{\hphantom{\mu}\nu}^{\mu}\right)^{\Psi} & =\frac{1}{4\kappa}\left[%
\partial^{\mu}\tilde{h}^{\alpha\beta}\partial_{\nu}\Psi_{\alpha\beta}-%
\partial^{\mu}\Psi^{\alpha\beta}\partial_{\nu}\Psi_{\alpha\beta}\right],
\label{eq:j-Psi-transl} \\
\left(j_{\hphantom{\mu}\nu}^{\mu}\right)^{\Phi} & =\frac{3}{2\kappa}%
\partial^{\mu}\Phi\partial_{\nu}\Phi,  \label{eq:j-Phi-transl} \\
\mathcal{L}A_{\nu}^{\mu}\left(x\right) & =0.  \label{eq:L-A-transl}
\end{align}

Inserting Eqs. (\ref{eq:j-h-tilde-transl})--(\ref{eq:L-A-transl}) into Eq. (%
\ref{eq:total-current}) leads to:
\begin{widetext}
\begin{equation}
t_{\mu\nu}\equiv\eta_{\mu\beta}\left\langle j_{\text{ \ }\nu}^{\beta}\right%
\rangle =\frac{c^{4}}{8\pi G}\left[\frac{1}{4}\left\langle \partial_{\mu}%
\tilde{h}^{\alpha\beta}\partial_{\nu}\tilde{h}_{\alpha\beta}\right\rangle -%
\frac{1}{4}\left\langle
\partial_{\mu}\Psi^{\alpha\beta}\partial_{\nu}\Psi_{\alpha\beta}\right%
\rangle +\frac{3}{2}\left\langle
\partial_{\mu}\Phi\partial_{\nu}\Phi\right\rangle \right].
\label{eq:t_munu-QG}
\end{equation}
\end{widetext}
This is the energy-momentum tensor of GW in quadratic gravity. As expected, $%
t_{\mu\nu}$ is obtained from the requirement of the Lagrangian symmetry
under spacetime translations.

The energy-momentum tensor $t_{\mu\nu}$ in Eq. (\ref{eq:t_munu-QG}) can be
compared to the $t_{\mu\nu}$ computed in Ref. \cite{Alves2022}. This
reference studies gravitational waves from inspiraling black holes in
quadratic gravity; the modified gravity theory therein is the same as the
one explored herein, thus one would expect consistency of the results. In
fact, Eq. (\ref{eq:t_munu-QG}) differs from Eq. (53) of Ref. \cite{Alves2022}
by the term $\left\langle \partial_{\mu}\tilde{h}^{\alpha\beta}\partial_{%
\nu}\Psi_{\alpha\beta}\right\rangle $, which is present in the latter.
However, it was shown in Ref. \cite{Alves2022} that this additional term
vanishes. Therefore, the geometrical approach to computing $t_{\mu\nu}$
adopted by Ref. \cite{Alves2022} is in complete accordance with the
field-theory approach adopted in this work. Here, there is a cancelation of the second term on the right-hand side (r.h.s.) of Eq. (\ref{eq:j-h-tilde-transl}) with the first term on the r.h.s. of Eq. (\ref{eq:j-Psi-transl}) on average after we neglect topological terms. The physical difference between the field-theoretical approach and the geometrical approach is precisely due to these topological terms. Because these terms are discarded in the bulk, the results of both approaches converge.

\subsubsection{The angular momentum of the GW in quadratic gravity}

The relativistic structure of quadratic gravity implies the existence of
rotational symmetry for the Lagrangian (\ref{eq:2ndOrderL}). This symmetry
is characterized by the matrix $\mathcal{R}_{ij}$ which acts on the space
coordinate $x^{i}$ to produce a rotation: $x^{\prime i}=\mathcal{R}_{%
\hphantom{i}j}^{i}x^{j}$. For infinitesimal transformations $\mathcal{R}%
_{ij} $ takes on the form\footnote{%
Recall that our metric signature is $\left(-,+,+,+\right)$.}
\begin{equation}
\mathcal{R}^{ij}=\delta^{ij}+\omega^{ij},  \label{eq:R-matrix}
\end{equation}
where $\omega^{ij}=-\omega^{ji}$ are the transformation parameters. Hence,
\begin{align}
x^{\prime i} &= x^{i}+\frac{1}{2}(\omega^{il}x^{l}-\omega^{ki}x^{k})\notag \\
 &= x^{i}+\sum_{k<l}\omega^{kl}\left(\delta^{ik}x^{l}-\delta^{il}x^{k}
\right)  \label{eq:rotations}
\end{align}
Comparing Eqs. (\ref{eq:rotations}) and (\ref{eq:coord-transf}):
\begin{equation}
A_{kl}^{i}=\delta^{i}_kx_{l}-\delta^{i}_lx_{k}\qquad\text{and}\qquad
A_{kl}^{0}=0.  \label{eq:A-rotations}
\end{equation}
A quick check of Eqs. (\ref{eq:field-transf}) and (\ref{eq:j-h-tilde-transl}%
)--(\ref{eq:j-Phi-transl}) shows that the rotation affects not only the
coordinates but also the fields $\tilde{h}_{\alpha\beta}$, $%
\Psi_{\alpha\beta}$ and $\Phi$. The field $\Phi$ is invariant under
rotations. The fields $\tilde{h}_{\alpha\beta}$ and $\Psi_{\alpha\beta}$ may
be decomposed into scalar, vector, and tensor irreducible quantities.

An arbitrary symmetric rank-2 tensor field $B_{\alpha\beta}$ decomposes into

\begin{itemize}
\item Two scalars: $B_{00}$ and $B=\eta^{ij}B_{ij}$ (spin-0 fields);

\item One space vector: $B_{0i}$ (spin-1 field);

\item One traceless space tensor: $\bar{B}_{ij}=B_{ij}-\frac{1}{3}\eta_{ij}B$
(spin-2 field).
\end{itemize}

Hence, by design, the components of the transformed field $%
B_{\alpha\beta}^{\prime}\left(x^{\prime}\right)$ under space rotation read:
\begin{equation}
B_{00}^{\prime }\left( x^{\prime }\right) =B_{00}\left( x\right) \text{ \
and \ }B^{\prime }\left( x^{\prime }\right) =B\left( x\right) ;
\label{eq:B-prime-scalar}
\end{equation}
\begin{align}
B_{0i}^{\prime }\left( x^{\prime }\right) & = \mathcal{R}_{i}^{\hphantom{i}%
j}B_{0j} \notag \\ 
& = B_{0i}+\sum_{k<l}\omega
^{kl}\left( \delta _{ik}B_{0l}-\delta _{il}B_{0k}\right) ;  \label{eq:B-prime-vector} 
\end{align}
\begin{align}
\bar{B}_{ij}^{\prime } & =\mathcal{R}_{i}^{\hphantom{i}k}\mathcal{R}_{j}^{%
\hphantom{i}l}\bar{B}_{kl} \notag \\
& = \bar{B}_{ij} + \sum_{k<l}\omega ^{kl}\left( \delta
_{kj}\bar{B}_{il}-\delta _{jl}\bar{B}_{ik}+\delta _{ik}\bar{B}_{lj}-\delta
_{il}\bar{B}_{kj}\right) .   \label{eq:B-prime-tensor}
\end{align}%
Eqs. (\ref{eq:B-prime-scalar})--(\ref{eq:B-prime-tensor}) allow us to
identify the quantities $F_{\alpha\beta,a}$ in (\ref{eq:field-transf})
describing field rotations. In fact, we have $a=\left[ik\right]$ with $%
\left(i,j\right)=1,2,3$, as follows:
\begin{align}
F_{00,kl}^{B} & =\eta^{ij}F_{ij,kl}^{B}=0,  \label{eq:F-scalar} \\
F_{0i,kl}^{B} & =\delta_{^{ik}}B_{0l}-\delta_{^{il}}B_{0k},
\label{eq:F-vector} \\
F_{ij,kl}^{B} & =\delta_{kj}\bar{B}_{il}-\delta_{jl}\bar{B}_{ik}+\delta_{ik}%
\bar{B}_{lj}-\delta_{il}\bar{B}_{kj}.  \label{eq:F-tensor}
\end{align}

The total angular momentum is defined as
\begin{equation}
J^{i}=\frac{1}{2c}\epsilon^{ilk}\int d^{3}x\left\langle
j_{kl}^{0}\right\rangle ,  \label{eq:total-angular-momentum}
\end{equation}
where $j_{kl}^{0}$ is the Noether conserved current---see Eq. (\ref%
{eq:conserved-current}). The current $j_{kl}^{\mu}$ can be split in two
sectors, viz. the orbital angular momentum density $l_{kl}^{\mu}$ and the
spin density $s_{kl}^{\mu}$:

\begin{equation}
j_{kl}^{\mu}=l_{kl}^{\mu}+s_{kl}^{\mu},  \label{eq:current-density-j}
\end{equation}
with
\begin{align}
l_{kl}^{\mu} & =\left(l_{kl}^{\mu}\right)^{\tilde{h}}+\left(l_{kl}^{\mu}%
\right)^{\Psi}+\left(l_{kl}^{\mu}\right)^{\Phi}-\mathcal{L}A_{kl}^{\mu},
\label{eq:current-density-l} \\
s_{kl}^{\mu} & =\left(s_{kl}^{\mu}\right)^{\tilde{h}}+\left(s_{kl}^{\mu}%
\right)^{\Psi}+\left(s_{kl}^{\mu}\right)^{\Phi}.
\label{eq:current-density-s}
\end{align}
Explicitly, the orbital sector is given by:
\begin{align}
\left(l_{kl}^{\mu}\right)^{\tilde{h}} & =\left(\frac{\partial\mathcal{L}}{%
\partial\partial_{\mu}\tilde{h}_{\alpha\beta}}-2\partial_{\nu}\frac{\partial%
\mathcal{L}}{\partial\partial_{\mu}\partial_{\nu}\tilde{h}_{\alpha\beta}}%
\right)A_{kl}^{\sigma}\partial_{\sigma}\tilde{h}_{\alpha\beta},
\label{eq:orbital-h-tilde} \\
\left(l_{kl}^{\mu}\right)^{\Psi} & =\left(\frac{\partial\mathcal{L}}{%
\partial\partial_{\mu}\Psi_{\alpha\beta}}-2\partial_{\nu}\frac{\partial%
\mathcal{L}}{\partial\partial_{\mu}\partial_{\nu}\Psi_{\alpha\beta}}%
\right)A_{kl}^{\sigma}\partial_{\sigma}\Psi_{\alpha\beta},
\label{eq:orbital-Psi} \\
\left(l_{kl}^{\mu}\right)^{\Phi} & =\left(\frac{\partial\mathcal{L}}{%
\partial\partial_{\mu}\Phi}-2\partial_{\nu}\frac{\partial\mathcal{L}}{%
\partial\partial_{\mu}\partial_{\nu}\Phi}\right)A_{kl}^{\sigma}\partial_{%
\sigma}\Phi,  \label{eq:orbital-Phi}
\end{align}
while the spin sector is computed from:
\begin{align}
\left(s_{kl}^{\mu}\right)^{\tilde{h}} & =-\left(\frac{\partial\mathcal{L}}{%
\partial\partial_{\mu}\tilde{h}_{\alpha\beta}}-2\partial_{\nu}\frac{\partial%
\mathcal{L}}{\partial\partial_{\mu}\partial_{\nu}\tilde{h}_{\alpha\beta}}%
\right)F_{\alpha\beta,kl},  \label{eq:spin-h-tilde} \\
\left(s_{kl}^{\mu}\right)^{\Psi} & =-\left(\frac{\partial\mathcal{L}}{%
\partial\partial_{\mu}\Psi_{\alpha\beta}}-2\partial_{\nu}\frac{\partial%
\mathcal{L}}{\partial\partial_{\mu}\partial_{\nu}\Psi_{\alpha\beta}}%
\right)F_{\alpha\beta,kl},  \label{eq:spin-Psi} \\
\left(s_{kl}^{\mu}\right)^{\Phi} & =-\left(\frac{\partial\mathcal{L}}{%
\partial\partial_{\mu}\Phi}-2\partial_{\nu}\frac{\partial\mathcal{L}}{%
\partial\partial_{\mu}\partial_{\nu}\Phi}\right)F_{\alpha\beta,kl}.
\label{eq:spin-Phi}
\end{align}

First, we compute the orbital angular momentum density. Henceforth, we
utilize the short-hand notation: $\partial_{t}\tilde{h}^{\alpha\beta}=\dot{h}%
^{\alpha\beta}$ (with the tilde suppressed for economy), $%
\partial_{t}\Psi^{\alpha\beta}=\dot{\Psi}^{\alpha\beta}$, $\partial_{t}\Phi=%
\dot{\Phi}$. Substituting Eqs. (\ref{eq:2ndOrderL}) and (\ref{eq:A-rotations}%
) into Eqs. (\ref{eq:orbital-h-tilde}), (\ref{eq:orbital-Phi}) and (\ref%
{eq:orbital-Psi}), results:
\begin{eqnarray}
l_{kl}^{0}=-\frac{1}{2\kappa c}[\dot{h}^{\alpha\beta}x^{[l}\partial^{k]}%
\tilde{h}_{\alpha\beta}-\dot{\Psi}^{\alpha\beta}x^{[l}\partial^{k]}\Psi_{%
\alpha\beta}\notag\\
+6\dot{\Phi}x^{[l}\partial^{k]}\Phi
+\left(CT\right)_{l}], \label{eq:l-zero}
\end{eqnarray}
where $\left(CT\right)_{l}$ stands for the crossed terms
\begin{equation}
\left(CT\right)_{l}\equiv\dot{h}^{\alpha\beta}x^{[l}\partial^{k]}\Psi_{%
\alpha\beta}-\dot{\Psi}^{\alpha\beta}x^{[l}\partial^{k]}\tilde{h}%
_{\alpha\beta}.  \label{eq:CT_l}
\end{equation}
By expliciting the average instructions, it is possible to demonstrate that $%
\left\langle \left(CT\right)_{l}\right\rangle $ is zero. In fact,
integrations by parts on average of the first term in Eq. (\ref{eq:CT_l}%
) lead to surface terms---which can be neglected ---and to the expression:
\begin{equation}
\left\langle \left(CT\right)_{l}\right\rangle =\left\langle \dot{\Psi}%
_{\alpha\beta}\left[\partial^{\lbrack k}\left(\tilde{h}^{\alpha\beta}x^{l]}%
\right)\right]\right\rangle -\left\langle \dot{\Psi}^{\alpha\beta}x^{[l}%
\partial^{k]}\tilde{h}_{\alpha\beta}\right\rangle =0,  \label{eq:CT_l-null}
\end{equation}
as anticipated.

The computation of the spin density is more evolved and requires the
decomposition of the tensor fields $\tilde{h}_{\alpha\beta}$ and $%
\Psi_{\alpha\beta}$ into irreducible quantities. After this task, one
evaluates the functional derivatives in (\ref{eq:spin-h-tilde}) and (\ref%
{eq:spin-Psi}), thus obtaining:
\begin{align}
\left( s_{kl}^{0}\right) ^{\tilde{h}} &=\frac{1}{2\kappa c}\left[ \frac{1}{2%
}\partial _{t}\bar{h}^{ij}+\frac{1}{2}\partial _{t}\bar{\Psi}^{ij}\right. 
\notag \\
&\left. -\alpha \partial _{t}\square \bar{h}^{ij}-\alpha \partial
_{t}\square \bar{\Psi}^{ij}\right] F_{ij,kl}^{\tilde{h}},  \label{eq:s-0-h-tilde(F)}
\end{align}%
and 
\begin{align}
\left( s_{kl}^{0}\right) ^{\Psi }& =\frac{1}{\kappa c}\left[ \partial
_{t}\Psi ^{0i}+\partial _{t}\tilde{h}^{0i}+6\gamma c\eta ^{ij}\partial
_{j}\square \Phi \right.   \notag \\
& \qquad \left. -2\alpha \partial _{t}\square \tilde{h}^{0i}-2\alpha \partial
_{t}\square \Psi ^{0i}\right] F_{0i,kl}^{\Psi }  \notag \\
& +\frac{1}{2\kappa c}\left[ \frac{1}{2}\partial _{t}\bar{\Psi}^{ij}+\frac{1%
}{2}\partial _{t}\bar{h}^{ij}\right.  \notag \\
& \qquad -\left. \alpha \partial _{t}\square \bar{h}^{ij}-\alpha \partial
_{t}\square \bar{\Psi}^{ij}\right] F_{ij,kl}^{\Psi },  \label{eq:s-0-Psi(F)}
\end{align}%
where the barred objects $\bar{h}^{ij} = \tilde{h}^{ij} - \frac{1}{3}\eta^{ij}\tilde{h}$ and $\bar{\Psi}^{ij} = \Psi^{ij} - \frac{1}{3}\eta^{ij}\Psi$ stand for the irreducible traceless representations of the spin-2 fields; $\tilde{h}$ and $\tilde{\Psi}$ denote the trace of their related tensors.
Eqs. (\ref{eq:s-0-h-tilde(F)}) and (\ref{eq:s-0-Psi(F)}) are further
simplified upon using the field equations in Eqs. (\ref{eq:FE-h-tilde}), (%
\ref{eq:FE-Psi}), (\ref{eq:FE-Phi}), the gauge conditions (\ref%
{eq:harmonic-TT-gauge}) and $\tilde{h}_{0\nu}=0$. (Under these conditions $%
\tilde{h}_{\mu\nu}$ is in the transverse-traceless gauge $\left(\text{TT}%
\right)$.) In effect,
\begin{align}
\left(s_{kl}^{0}\right)^{\tilde{h}} & =\frac{1}{4\kappa c}\left(\partial_{t}%
\tilde{h}^{ij}-\partial_{t}\Psi^{ij}\right)F_{ij,kl}^{\tilde{h}}
\label{eq:s-0-h-tilde} \\
\left(s_{kl}^{0}\right)^{\Psi} & =-\frac{1}{\kappa c}\left[%
\partial_{t}\Psi^{0i}+2c\partial^{i}\Phi\right]F_{0i,kl}^{\Psi} \notag \\
& \quad \, -\frac{1}{4\kappa c}\left[\partial_{t}\Psi^{ij}-\partial_{t}\tilde{h}^{ij}\right]%
F_{ij,kl}^{\Psi}.  \label{eq:s-0-Psi}
\end{align}
At last, Eqs. (\ref{eq:F-vector}) and (\ref{eq:F-tensor}) are plugged into
Eqs. (\ref{eq:s-0-h-tilde}) and (\ref{eq:s-0-Psi}). In the face of Eq. (\ref%
{eq:current-density-s}), we get to
\begin{align}
s_{kl}^{0} &=\frac{1}{\kappa c}\left[ h_{a[l}^{TT}\dot{h}_{k]a}^{TT}-\Psi
_{a[l}\dot{\Psi}_{k]a}\right.  \notag \\
&\left. +2\Psi _{0[l}\dot{\Psi}_{k]0}+\left( CT\right) _{s_{1}}-4c\left(
CT\right) _{s_{2}}\right]  \label{eq:s-zero}
\end{align}
where
\begin{align}
\left(CT\right)_{s_{1}} & \equiv\Psi_{a[l}\dot{h}_{k]a}^{TT}-h_{a[l}^{TT}%
\dot{\Psi}_{k]a},  \label{eq:CT_s1} \\
\left(CT\right)_{s_{2}} & \equiv\Psi_{0[l}\partial_{k]}\Phi\text{.}
\label{eq:CT_s2}
\end{align}   
The crossed terms $\left(CT\right)_{s_{1}}$ and $\left(CT\right)_{s_{2}}$
are also null. Consider the term in Eq. (\ref{eq:CT_s1}). Integrating by
parts with respect to time and neglecting the surface term, while taking the
average, gives:
\begin{equation}
\left\langle \left(CT\right)_{s_{1}}\right\rangle =\left\langle \Psi_{a[l}%
\dot{h}_{k]a}^{TT}\right\rangle +\left\langle \dot{h}_{a[l}^{TT}\Psi_{k]a}%
\right\rangle =0.  \label{eq:CT_s1-null}
\end{equation}
For computing the term $\left(CT\right)_{s_{2}}$, we make use of the field
equations (\ref{eq:FE-Psi}) and (\ref{eq:FE-Phi}):
\begin{eqnarray}
m_{\Phi}^{2}\Phi & =\square\Phi\Rightarrow
m_{\Phi}^{2}\Psi_{0l}\partial_{k}\Phi=\Psi_{0l}\partial_{k}\square\Phi,
\label{eq:CT_s2-FE-Phi} \\
m_{\Psi}^{2}\Psi_{0l} & =\square\Psi_{0l}\Rightarrow
m_{\Psi}^{2}\Psi_{0l}\partial_{k}\Phi=\square\Psi_{0l}\partial_{k}\Phi.
\label{eq:CT_s2-FE-Psi}
\end{eqnarray}
Subtracting (\ref{eq:CT_s2-FE-Phi}) and (\ref{eq:CT_s2-FE-Psi}) and taking the average yields
\begin{align}
\left(m_{\Phi}^{2}-m_{\Psi}^{2}\right)\left\langle
\Psi_{0l}\partial_{k}\Phi\right\rangle & =\left\langle
\Psi_{0l}\partial_{k}\square\Phi\right\rangle -\left\langle
\square\Psi_{0l}\partial_{k}\Phi\right\rangle =0\Rightarrow  \notag \\
\Rightarrow\left\langle \Psi_{0l}\partial_{k}\Phi\right\rangle & =0\qquad%
\text{if}\qquad m_{\Psi}\neq m_{\Phi}\text{.}  \label{eq:CT_s2-null}
\end{align}
Analogously, $\left\langle \Psi_{0k}\partial_{l}\Phi\right\rangle =0$. The
last two results then demand $\left\langle
\left(CT\right)_{s_{2}}\right\rangle =0$.

After these necessary intermediate calculations, we substitute Eqs. (\ref%
{eq:current-density-j}), (\ref{eq:l-zero}), and (\ref{eq:s-zero}) into Eq. (%
\ref{eq:total-angular-momentum}) to obtain the final form of the total
angular momentum for GW in quadratic gravity:
\begin{align}
J^{i} & = \frac{c^{2}}{32\pi G}\int d^{3}x\left[ \epsilon ^{ikl}\left(
-\left\langle \dot{h}_{ab}^{TT}x_{k}\partial _{l}\tilde{h}%
_{ab}^{TT}\right\rangle \right. \right. \notag \\
& +2\left\langle h_{ak}^{TT}\dot{h}_{al}^{TT}\right \rangle  +\left\langle \dot{\Psi}^{\alpha \beta }x_{k}\partial
_{l}\Psi _{\alpha \beta }\right\rangle -2\left\langle \Psi _{ak}\dot{\Psi}_{al}\right\rangle  \notag  \\
& \left. \left. +4\left\langle \Psi _{0k}\dot{\Psi}_{0l}\right\rangle
-6\left\langle \dot{\Phi}x_{k}\partial _{l}\Phi \right\rangle \right) \right]
.   \label{eq:total-angular-momentum-QG}
\end{align}%
The first two terms of the integrand reproduce the result known from general
relativity \cite{Maggiore2007}. The last four are extra terms related to the
spin-2 and spin-0 massive fields.
\subsubsection{Energy and angular momentum radiated}
We are interested in computing the energy and angular momentum carried away
from the source by the gravitational waves. The gravitational radiation at a
sufficiently distant point from the source (radiation zone) is structurally
given by Refs. \cite{Alves2022,Vilhena2021}:
\begin{equation}
F_{i}=\frac{1}{r}\sum_{n}f_{i}^{\left(n\right)}\left(\omega_{n}\left[t-\frac{%
r}{v_{p}^{\left(n\right)}}\right]\right).  \label{eq:F_i}
\end{equation}
Here $n$ labels the mode of frequency $\omega_{n}$; $v_{p}^{\left(n\right)}$
is the speed of this mode. Then,
\begin{align}
\tilde{h}_{\alpha\beta}^{\left(n\right)} & \rightarrow
v_{p}^{\left(n\right)}=c,   \\
\Psi_{\alpha\beta}^{\left(n\right)} & \rightarrow v_{p}^{\left(n\right)}=%
\frac{c}{\sqrt{1-\left(\frac{m_{\Psi}c}{\omega_{n}}\right)^{2}}},
\label{eq:v_p(n)} \\
\Phi^{\left(n\right)} & \rightarrow v_{p}^{\left(n\right)}=\frac{c}{\sqrt{%
1-\left(\frac{m_{\Phi}c}{\omega_{n}}\right)^{2}}}, 
\end{align}
where $\omega_{n}>m_{X}c$ with $X=\Psi,\Phi$.

By invoking the conservation of the energy-momentum tensor (\ref%
{eq:t_munu-QG}), we show that the total energy $E$ emitted by the source is:
\begin{align}
\frac{dE}{dt} &=-c\int r^{2}d\Omega t_{0r} \notag \\
&=-\frac{c^{5}r^{2}}{32\pi G}\sum_{m,n}\int d\Omega \left[ \left\langle
\partial _{0}\tilde{h}_{\left( n\right) }^{\alpha \beta }\partial _{r}\tilde{%
h}_{\alpha \beta }^{\left( m\right) }\right\rangle \right.   \notag \\
&\left. -\left\langle \partial _{0}\Psi _{\left( n\right) }^{\alpha \beta
}\partial _{r}\Psi _{\alpha \beta }^{\left( m\right) }\right\rangle
+6\left\langle \partial _{0}\Phi _{\left( n\right) }\partial _{r}\Phi
^{\left( m\right) }\right\rangle \right] .  \label{eq:dEdt}
\end{align}
Each frequency $\omega_{n}$ is characterized by the Fourier mode
\begin{equation}
f_{i}^{\left(n\right)}\sim\cos\left[\omega_{n}\left(t-\frac{r}{%
v_{p}^{\left(n\right)}}\right)\right]_{i}\text{.} \label{eq:f_i(n)}
\end{equation}
The averaging process combined with the orthogonality properties of the
Fourier series leads to
\begin{align}
\sum_{m,n}\left\langle
\partial_{0}f_{\left(n\right)}^{i}\partial_{r}f_{i}^{\left(m\right)}\right \rangle & = \sum_{m,n}\left\langle
\partial_{0}f_{\left(n\right)}^{i}\partial_{r}f_{i}^{\left(m\right)}\right \rangle \delta_{\left(m\right)}^{\left(n\right)} \notag \\
&=\sum_{n}\left\langle
\partial_{0}f_{\left(n\right)}^{i}\partial_{r}f_{i}^{\left(n\right)}\right%
\rangle.  \label{eq:ortodonality-df_i}
\end{align}
Moreover, at great distances from the source, the observer perceives the
wavefront as a plane wave. For this reason, in the radiation zone, the
radial derivatives can be approximated by:
\begin{equation}
\partial_{r}f_{i}^{\left(n\right)}\approx-\frac{1}{v_{p}^{\left(n\right)}}%
\dot{f}_{i}^{\left(n\right)}.  \label{eq:df_idr-approx}
\end{equation}
Given these facts,
\begin{align}
\frac{dE}{dt} &=\frac{c^{3}r^{2}}{32\pi G}\sum_{n}\int d\Omega \left[
\left\langle \dot{h}_{\left( n\right) }^{\alpha \beta }\dot{h}_{\alpha \beta
}^{\left( n\right) }\right\rangle \right.  \notag \\
&\left. -q_{n}^{\Psi }\left\langle \dot{\Psi}_{\left( n\right) }^{\alpha
\beta }\dot{\Psi}_{\alpha \beta }^{\left( n\right) }\right\rangle \Theta
_{n}^{\Psi }+6q_{n}^{\Phi }\left\langle \dot{\Phi}_{\left( n\right)
}^{2}\right\rangle \Theta _{n}^{\Phi }\right] ,  \label{eq:energy-loss}
\end{align}%
where
\begin{equation}
q_{n}^{\Psi}\equiv\sqrt{1-\left(\frac{m_{\Psi}c}{\omega_{n}}\right)^{2}}, \quad \rm{and} \quad 
q_{n}^{\Phi}\equiv\sqrt{1-\left(\frac{m_{\Phi}c}{\omega_{n}}%
\right)^{2}},  \label{eq:q_n}
\end{equation}
with
\begin{equation}
\Theta_{n}^{\Psi}\equiv\Theta\left(\omega_{n}-m_{\Psi}c\right), \quad \rm{and} \quad
\Theta_{n}^{\Phi}\equiv\Theta\left(\omega_{n}-m_{\Phi}c\right),
\label{eq:Heaviside}
\end{equation}
being Heaviside functions which appear to inform that the massive modes are
propagating modes only for $\omega_{n}>m_{X}c$. The negative sign in the
second term of Eq. (\ref{eq:energy-loss}) indicates the occurrence of an
Ostrogradsky instability; this fact is traced back to the term $\alpha
C_{\mu\nu\alpha\beta}C^{\mu\nu\alpha\beta}$ in the action (\ref{eq:S}).

The equation for the radiated angular momentum can be obtained from the
conservation of the current $j_{kl}^{\mu}$ given by Eq. (\ref%
{eq:current-density-j}). However, it is simpler to perform a heuristic
construction based on the total angular momentum \cite{Maggiore2007}. In
fact, the following expression for the angular momentum density $j^{i}/c$ is
inferred from the structure of Eq. (\ref{eq:total-angular-momentum-QG}):
\begin{equation}
\frac{j^{i}}{c}=\sum_{n}\left(\frac{j_{h}^{i\left(n\right)}}{c}+\frac{%
j_{\Psi}^{i\left(n\right)}}{c}+\frac{j_{\Phi}^{i\left(n\right)}}{c}\right)
\label{eq:j-i}
\end{equation}
where
\begin{align}
\frac{j_{h}^{i\left( n\right) }}{c}& =\frac{r^{2}c^{2}}{32\pi G}\epsilon
^{ikl}\left[ -\left\langle \dot{h}_{ab}^{\left( n\right) }x_{k}\partial
_{l}h_{ab}^{\left( n\right) }\right\rangle +2\left\langle h_{ak}^{\left(
n\right) }\dot{h}_{al}^{\left( n\right) }\right\rangle \right] ,
\label{eq:j-i-h} \\
\frac{j_{\Psi }^{i\left( n\right) }}{c}& =\frac{r^{2}c^{2}}{32\pi G}\epsilon
^{ikl}\left[ \left\langle \dot{\Psi}_{\left( n\right) }^{\alpha \beta
}x_{k}\partial _{l}\Psi _{\alpha \beta }^{\left( n\right) }\right\rangle \right.  \notag \\
& \left. 
-2\left\langle \Psi _{ak}^{\left( n\right) }\dot{\Psi}_{al}^{\left( n\right)
}\right\rangle +4\left\langle \Psi _{0k}^{\left( n\right) }\dot{\Psi}_{0l}^{\left(
n\right) }\right\rangle \right] ,   \label{eq:j-i-Psi} \\
\frac{j_{\Phi }^{i\left( n\right) }}{c}& =\frac{r^{2}c^{2}}{32\pi G}\epsilon
^{ikl}\left[ -6\left\langle \dot{\Phi}_{\left( n\right) }x_{k}\partial
_{l}\Phi ^{\left( n\right) }\right\rangle \right] .  \label{eq:j-i-Phi}
\end{align}
Here, we have suppressed the labeling for the TT gauge for economic reason,
i.e. $h_{ab}=h_{ab}^{TT}$.

The radiated gravitational wave of frequency $\omega_{n}$ is built from the
various modes $h_{\alpha\beta}^{\left(n\right)}$, $\Psi_{\alpha\beta}^{%
\left(n\right)}$, and $\Phi^{\left(n\right)}$ that propagate with group
velocities $v_{g}^{\left(n\right)}$:\footnote{%
Section VI of Ref. \cite{Alves2022} provides details on the group velocity of
massive modes.}
\begin{eqnarray}
\tilde{h}_{\alpha\beta}^{\left(n\right)} & \rightarrow
v_{g}^{\left(n\right)}=c,  \notag \\
\Psi_{\alpha\beta}^{\left(n\right)} & \rightarrow
v_{g}^{\left(n\right)}=cq_{n}^{\Psi},  \label{eq:v_g(n)} \\
\Phi^{\left(n\right)} & \rightarrow v_{g}^{\left(n\right)}=cq_{n}^{\Phi}.
\notag
\end{eqnarray}
Consider a wavefront sector of a specific mode, at the time $t$ and radial
distance $r$ from the source, subtending a solid angle $d\Omega$. After an
infinitesimal time interval $dt$, this wavefront sector has swept a volume $%
d^{3}x=r^{2}drd\Omega=r^{2}\left(v_{g}^{\left(n\right)}dt\right)d\Omega$.
Since the angular momentum density of each mode is given by $%
j^{i\left(n\right)}/c$, the angular momentum carried by this mode will be $%
dJ^{i\left(n\right)}=\left(j^{i\left(n\right)}/c\right)r^{2}\left(v_{g}^{%
\left(n\right)}dt\right)d\Omega$. Therefore, the radiated angular momentum
is given by:
\begin{equation}
\frac{dJ^{i}}{dt}=\sum_{n}\int
d\Omega\left(j_{h}^{i\left(n\right)}+q_{n}^{\Psi}j_{\Psi}^{i\left(n\right)}%
\Theta_{n}^{\Psi}+q_{n}^{\Phi}j_{\Phi}^{i\left(n\right)}\Theta_{n}^{\Phi}%
\right),  \label{eq:angular-momentum-loss}
\end{equation}
where $q_{n}$ and $\Theta_{n}$ are the quantities in Eqs. (\ref{eq:q_n}) and
(\ref{eq:Heaviside}).

Eqs. (\ref{eq:energy-loss}) and (\ref{eq:angular-momentum-loss}) are the
main results of this section.

\section{Quadrupole approximation\label{sec:Quadrupole-approx}}

In this section, we present the general structure of the propagation modes
of quadratic gravity in the quadrupole approximation. In particular, we are
interested in describing the combination of the various modes and in
addressing the Ostrogradsky instability exhibited by the theory.

The field equations (\ref{eq:FE-h-tilde}), (\ref{eq:FE-Psi}), and (\ref%
{eq:FE-Phi}) in the presence of matter read
\begin{eqnarray}
\square\tilde{h}_{\mu\nu} & =-2\kappa T_{\mu\nu},  \label{eq:FE-h-matter} \\
\left(\square-m_{\Psi}^{2}\right)\Psi_{\mu\nu} & =2\kappa T_{\mu\nu},
\label{eq:FE-Psi-matter} \\
\left(\square-m_{\Phi}^{2}\right)\Phi & =-\frac{\kappa T}{3},
\label{eq:FE-Phi-matter}
\end{eqnarray}
where $T_{\mu\nu}$ is the matter energy momentum tensor and $T$ is its trace.

The solutions to the above equations in a region devoid of sources and in
the quadrupole approximation are \cite{Maggiore2007,Alves2022}:
\begin{align}
\tilde{h}_{ij}^{TT}\left( \mathbf{x},t\right) & =\frac{\kappa }{4\pi r}%
\Lambda _{ij,kl}\left( \mathbf{\hat{n}}\right) \ddot{M}^{kl}\left(
t_{r}\right) ,  \label{eq:h_ij} \\
\Psi _{ij}\left( \mathbf{x},t\right) & =-\frac{\kappa }{4\pi r}\ddot{M}_{ij}\left( t_{r}\right) \notag  \\
& +\frac{\kappa }{4\pi }m_{\Psi }\int_{0}^{\infty }d\bar{t}_{r}F_{\Psi
}\left( \bar{t}_{r}\right) \ddot{M}_{ij}\left( \zeta \right) ,    \label{eq:Psi_ij} \\
\Phi ^{Q}\left( \mathbf{x},t\right) & =\frac{\kappa }{24\pi r}n_{i}n_{j}%
\mathcal{\ddot{M}}^{ij}\left( t_{r}\right)  \notag \\
& -\frac{\kappa }{24\pi }m_{\Phi }\int_{0}^{\infty }d\bar{t}_{r}F_{\Phi
}\left( \bar{t}_{r}\right) n_{i}n_{j}\mathcal{\ddot{M}}^{ij}\left( \zeta
\right) .   \label{eq:Phi_Q}
\end{align}%
Here $\Lambda_{ij,kl}$ is the projection tensor in the TT gauge \cite{Maggiore2007}, $t_{r}=t-%
\frac{r}{c}$ is the retarded time, $\zeta\equiv t_{r}-\bar{t}_{r}$, and $%
n_{i}$ points in the direction of propagation of the wave. The quadrupole
mass moments are computed from the expressions
\begin{eqnarray}
M^{ij} & =\frac{1}{c^{2}}\int d^{3}xT^{00}\left(t,\mathbf{x}%
\right)x^{i}x^{j},  \label{eq:mass-moment(T00)} \\
\mathcal{M}^{ij} & =\frac{1}{c^{2}}\int d^{3}\mathbf{x}T\left(\mathbf{x}%
,t\right)x^{i}x^{j},  \label{eq:mass-moment(T)}
\end{eqnarray}
where
\begin{equation}
F_{X}\left(\bar{t}_{r}\right)=\frac{J_{1}\left(m_{X}c\sqrt{2\bar{t}_{r}}%
\sqrt{\frac{\bar{t}_{r}}{2}+\frac{r}{c}}\right)}{\sqrt{2\bar{t}_{r}}\sqrt{%
\frac{\bar{t}_{r}}{2}+\frac{r}{c}}},  \label{eq:F(t_r)}
\end{equation}
with the Bessel function of first type $J_{1}$ and, again, $X=\Psi$,$\Phi$.
We refer the interested reader to Ref. \cite{Alves2022} for further details.

In general, the scalar mode is able to produce emission of the dipole type.
However, this mode is null for non-relativistic systems where the total
momentum is conserved. In this situation, the dominant term of $\Phi$ is the
quadrupole term and
\begin{equation}
T=\eta_{\mu\nu}T^{\mu\nu}=-T^{00}\Rightarrow\mathcal{M}^{ij}=-M^{ij}\text{.}
\label{eq:mass-moment}
\end{equation}

\subsection{GWs propagation in a generic direction\label%
{subsec:general-direction}}

The maximum number of propagating degrees of freedom is six for a
metric-compatible four-dimensional gravity theory invariant under
diffeomorphisms. This number reduces to two in GR due to the residual gauge
freedom. In the context of any metric theory of gravity, there are at most six degrees of freedom. The reason for this is the covariance of the theory, which allows us to choose a coordinate system eliminating four of the ten degrees of freedom of the metric tensor. In the particular case of quadratic gravity, the six degrees of
freedom are: five carried by $\tilde{h}_{ij}$ and $\Psi_{ij}$, and one related to the scalar $\Phi$--cf., the discussion at the end of this section. These six degrees of freedom are
engendered by six components of the quadrupole mass momentum $M_{ij}$
(disregarding the gravitational back-reaction). In the non-relativistic
regime and under the quadrupole approximation, the spin-0 part is produced
by the single component $n_{i}n_{j}\mathcal{\ddot{M}}^{ij}=-n_{i}n_{j}\ddot{M%
}^{ij}$. The spin-2 contribution is obtained from $\ddot{M}^{ij}$ with a
constraint coming from the trace of $\ddot{M}^{ij}$.

In order to better understand the structure of the theory in this regard,
let us begin the analysis by taking a wave propagating in the $z$-direction,
i.e. $n_{i}^{\prime}=\left(0,0,1\right)$.

The (massless) spin-2 mode is equivalent to that of general relativity.
Hence,
\begin{equation}
h_{+}=\frac{\tilde{h}_{11}-\tilde{h}_{22}}{2}\qquad\text{and}\qquad
h_{\times}=\tilde{h}_{12};  \label{eq:h-polarization}
\end{equation}
the $h_{+}$ and $h_{\times}$ are the two degrees of freedom of $\tilde{h}%
_{\mu\nu}$. Thus, comparison with Eqs. (\ref{eq:h_ij}) and (\ref%
{eq:h-polarization}) yields:
\begin{align}
h_{+}\left(t,\mathbf{\hat{z}}\right) & =\frac{\kappa}{8\pi r}\left(\ddot{M}%
_{11}^{\prime}-\ddot{M}_{22}^{\prime}\right),  \label{eq:h_plus(z)} \\
h_{+}\left(t,\mathbf{\hat{z}}\right) & =\frac{\kappa}{4\pi r}\ddot{M}%
_{12}^{\prime}.  \label{eq:h_cross(z)}
\end{align}

The massive spin-2 mode has five degrees of freedom. Aiming for a
description similar to that in GR, we define \cite{Alves2022}:
\begin{eqnarray}
\Psi_{+} =\frac{\Psi_{11}-\Psi_{22}}{2}, \quad \Psi_{\times}=\Psi_{12},  \notag \\
\Psi_{B}  =\Psi_{13} , \quad \Psi_{C}=\Psi_{23} \, \quad \text{and}%
\quad \Psi_{D}=\Psi_{33},  \label{eq:Psi-components}
\end{eqnarray}
where $\Psi_{+}$ and $\Psi_{\times}$ are the transverse propagation modes,
while $\Psi_{B}$, $\Psi_{C}$, and $\Psi_{D}$ are the longitudinal
propagation modes. The integral operator
\begin{equation}
\hat{O}_{X}M_{ij}\equiv rm_{X}\int_{0}^{\infty}d\bar{t}_{r}F_{X}\left(\bar{t}%
_{r}\right)M_{ij}\left(\zeta\right)  \label{eq:O-hat}
\end{equation}
and Eqs. (\ref{eq:Psi_ij}) and (\ref{eq:Psi-components}) lead to:

\begin{align}
\Psi_{+}\left(t,\mathbf{\hat{z}}\right) & =-\frac{\kappa}{8\pi r}\left[1-%
\hat{O}_{\Psi}\right]\left(\ddot{M}_{11}^{\prime}-\ddot{M}%
_{22}^{\prime}\right),  \label{eq:Psi_plus(z)} \\
\Psi_{\times}\left(t,\mathbf{\hat{z}}\right) & =-\frac{\kappa}{4\pi r}\left[%
1-\hat{O}_{\Psi}\right]\ddot{M}_{12}^{\prime},  \label{eq:Psi_cross(z)} \\
\Psi_{B}\left(t,\mathbf{\hat{z}}\right) & =-\frac{\kappa}{4\pi r}\left[1-%
\hat{O}_{\Psi}\right]\ddot{M}_{13}^{\prime},  \label{eq:Psi_B(z)} \\
\Psi_{C}\left(t,\mathbf{\hat{z}}\right) & =-\frac{\kappa}{4\pi r}\left[1-%
\hat{O}_{\Psi}\right]\ddot{M}_{23}^{\prime},  \label{eq:Psi_C(z)} \\
\Psi_{D}\left(t,\mathbf{\hat{z}}\right) & =-\frac{\kappa}{4\pi r}\left[1-%
\hat{O}_{\Psi}\right]\ddot{M}_{33}^{\prime}.  \label{eq:Psi_D(z)}
\end{align}

The spin-0 mode is built following an analogous procedure. The result is:
\begin{equation}
\Phi\left(t,\mathbf{\hat{z}}\right)=-\frac{\kappa}{24\pi r}\left[1-\hat{O}%
_{\Phi}\right]\ddot{M}_{33}^{\prime}.  \label{eq:Phi(z)}
\end{equation}

Next, we generalize the expressions for the mode for a generic propagation
direction oriented by the unit vector
\begin{equation}
n_{i}=\left(\sin\theta\sin\phi,\sin\theta\cos\phi,\cos\theta\right),
\label{eq:n_i}
\end{equation}
where $\theta$ is the polar angle and $\phi$ is the azimuth angle \cite%
{Maggiore2007}. The rotation matrix connecting the directions $n_{i}$ and $
n_{i}^{\prime}$ is:
\begin{eqnarray}
R =\left(%
\begin{array}{ccc}
\cos\phi & \cos\theta\sin\phi & \sin\theta\sin\phi \\
-\sin\phi & \cos\theta\cos\phi & \cos\phi\sin\theta \\
0 & -\sin\theta & \cos\theta%
\end{array}%
\right),  \label{eq:R-matriz}\\ \notag
\end{eqnarray}
since $n_{i}=R_{ij}n_{j}^{\prime}$. Moreover, the mass moments in an
arbitrary direction by acting the rotation matrix onto $M_{ij}$:
\begin{equation*}
M_{ij} = R_{ia}R_{jb}M_{ab}^{\prime}\Rightarrow 
\mathbf{M}  =\mathbf{RM}^{\prime} \mathbf{R}^{T}
\end{equation*}
i.e.,
\begin{equation}
\mathbf{M}^{\prime} = \mathbf{R}^{T}\mathbf{MR.}
\label{eq:M-general}
\end{equation}
Eq. (\ref{eq:M-general}) teaches us how to write the propagating modes of
Eqs. (\ref{eq:h_plus(z)}), (\ref{eq:h_cross(z)}), (\ref{eq:Psi_plus(z)}), (%
\ref{eq:Psi_cross(z)}), (\ref{eq:Psi_B(z)}), (\ref{eq:Psi_C(z)}), (\ref%
{eq:Psi_D(z)}), and (\ref{eq:Phi(z)}) in a generic direction:
\begin{itemize}
\item Massless spin-2 modes:
\begin{align}
h_{+} & \left( t,\theta ,\phi \right) =\frac{\kappa }{8\pi r}\left[ \ddot{M}%
_{11}\left( \cos ^{2}\phi -\cos ^{2}\theta \sin ^{2}\phi \right) \right.  
\notag \\
& +\ddot{M}_{22}\left( \sin ^{2}\phi -\cos ^{2}\theta \cos ^{2}\phi \right) -%
\ddot{M}_{33}\sin ^{2}\theta   \notag \\
& -\ddot{M}_{12}\left( 1+\cos ^{2}\theta \right) \sin \left( 2\phi \right) +\ddot{M}_{13}\sin \left( 2\theta \right) \sin \phi   \notag \\
& \left. +\ddot{M}_{23}\sin \left( 2\theta \right) \cos \phi \right],
\label{eq:h_plus}
\end{align}
\begin{align}
h_{\times }\left( t,\theta ,\phi \right) & =\frac{\kappa }{8\pi r}\left[
\left( \ddot{M}_{11}-\ddot{M}_{22}\right) \cos \theta \sin \left( 2\phi
\right) \right.   \notag \\
& +2\ddot{M}_{12}\cos \theta \cos \left( 2\phi \right) -2\ddot{M}_{13}\sin
\theta \cos \phi   \notag \\
& \left. +2\ddot{M}_{23}\sin \theta \sin \phi \right] .  \label{eq:h_cross}
\end{align}
\item Massive spin-2 modes: 
\begin{align}
\Psi _{+} & \left( t,\theta ,\phi \right) =-\frac{\kappa }{8\pi r}\left[ 1-%
\hat{O}_{\Psi }\right] \left[ \ddot{M}_{11}\cos ^{2}\phi \right.   \notag \\
& -\ddot{M}_{11}\cos ^{2}\theta \sin ^{2}\phi +\ddot{M}_{22}\sin ^{2}\phi  -\ddot{M}_{22}\cos ^{2}\theta \cos ^{2}\phi 
\notag \\
& -\ddot{M}_{33}\sin ^{2}\theta -\ddot{M}_{12}\left( 1+\cos ^{2}\theta \right) \sin \left( 2\phi \right)  
\notag \\
& \left. +\ddot{M}_{13}\sin \left( 2\theta \right) \sin \phi +\ddot{M}%
_{23}\sin \left( 2\theta \right) \cos \phi \right] ,  \label{eq:Psi_plus}
\end{align}
\begin{align}
\Psi _{\times }\left( t,\theta ,\phi \right) & =-\frac{\kappa }{8\pi r}\left[
1-\hat{O}_{\Psi }\right] \left[ \ddot{M}_{11}\cos \theta \sin \left( 2\phi
\right) \right.   \notag \\
& -\ddot{M}_{22}\cos \theta \sin \left( 2\phi \right) +2\ddot{M}_{12}\cos
\theta \cos \left( 2\phi \right)   \notag \\
& \left. -2\ddot{M}_{13}\sin \theta \cos \phi  +2\ddot{M}_{23}\sin \theta \sin \phi \right] ,  \label{eq:Psi_cross}
\end{align}
\begin{align}
\Psi _{B} \left( t,\theta ,\phi \right) & =-\frac{\kappa }{8\pi r}\left[ 1-%
\hat{O}_{\Psi }\right] \left[ \ddot{M}_{11} \sin
\theta \sin \left( 2\phi \right) \right.   \notag \\
& -\ddot{M}_{22} \sin \theta \sin \left( 2\phi \right) +2\ddot{M}_{12}\sin \theta \cos \left( 2\phi \right)   \notag   \\
& \left. +2\ddot{M}_{13}\cos
\theta \cos \phi -2\ddot{M}_{23}\cos \theta \sin \phi \right] , \label{eq:Psi_B}
\end{align}%
\begin{align}
\Psi _{C}\left( t,\theta ,\phi \right) & =-\frac{\kappa }{8\pi r}\left[ 1-%
\hat{O}_{\Psi }\right] \left[ \ddot{M}_{11}\sin ^{2}\phi \sin \left( 2\theta
\right) \right.   \notag \\
& +\ddot{M}_{22}\cos ^{2}\phi \sin \left( 2\theta \right)  -\ddot{M}_{33}\sin \left( 2\theta \right)  \notag \\
& +\ddot{M}_{12}\sin \left(
2\theta \right) \sin \left( 2\phi \right) +2\ddot{M}_{13}\cos \left( 2\theta \right) \sin \phi   \notag \\
& \left. +2\ddot{M}%
_{23}\cos \left( 2\theta \right) \cos \phi \right] ,  \label{eq:Psi_C}
\end{align}
\begin{align}
\Psi _{D}\left( t,\theta ,\phi \right) & =-\frac{\kappa }{4\pi r}\left[ 1-\hat{O}_{\Psi }\right] \left[ 
\ddot{M}_{11}\sin ^{2}\phi \sin ^{2}\theta \right.   \notag \\
& +\ddot{M}_{22}\cos ^{2}\phi \sin ^{2}\theta +\ddot{M}_{33}\cos ^{2}\theta 
\notag \\
& +\ddot{M}_{12}\sin ^{2}\theta \sin \left( 2\phi \right) +\ddot{M}_{13}\sin
\left( 2\theta \right) \sin \phi   \notag \\
& \left. +\ddot{M}_{23}\sin \left( 2\theta \right) \cos \phi \right] .
\label{eq:Psi_D}
\end{align}

\item Massive spin-$0$ mode: 
\begin{align}
\Phi \left( t,\theta ,\phi \right)& =-\frac{\kappa }{24\pi r}\left[ 1-\hat{O%
}_{\Phi }\right] \left[ \ddot{M}_{11}\sin ^{2}\phi \sin ^{2}\theta \right.  
\notag \\
& +\ddot{M}_{22}\cos ^{2}\phi \sin ^{2}\theta +\ddot{M}_{33}\cos ^{2}\theta 
\notag \\
& +\ddot{M}_{12}\sin ^{2}\theta \sin \left( 2\phi \right) +\ddot{M}_{13}\sin \left( 2\theta \right) \sin \phi   \notag \\
& \left. +\ddot{M}_{23}\sin \left( 2\theta \right) \cos \phi \right] .  \label{eq:Phi}
\end{align}
\end{itemize}
The gravitational wave to be detected in a given direction $%
\left(\theta,\phi\right)$ is the result of an interference process involving
all the modes above. However, the form of the equations (\ref{eq:h_plus})--(%
\ref{eq:Phi}) shows that the interference process occurs as combinations of $%
\left\{ h_{+},\Psi_{+}\right\} $, $\left\{ h_{\times},\Psi_{\times}\right\} $%
, $\Psi_{B}$, $\Psi_{C}$, $\Psi_{D}$, and $\Phi$. This is corroborated by
the analysis of the particular case of a wave of frequency $\omega_{n}$
propagating along the $z$-direction $\left(\theta=\phi=0\right)$. In this
case, Eq. (\ref{eq:h-bar(h-tilde,Psi,Phi)}) for $\bar{h}_{ij}$ decomposes
into
\begin{align}
& \bar{h}_{ij}^{\left(n\right)} = \left(%
\begin{array}{ccc}
h_{+}^{\left(n\right)}+\Psi_{+}^{\left(n\right)} & h_{\times}^{\left(n%
\right)}+\Psi_{\times}^{\left(n\right)} & 0 \\
h_{\times}^{\left(n\right)}+\Psi_{\times}^{\left(n\right)} &
-\left(h_{+}^{\left(n\right)}+\Psi_{+}^{\left(n\right)}\right) & 0 \\
0 & 0 & 0%
\end{array}%
\right) \notag
\\
&+\left(%
\begin{array}{ccc}
\Phi^{\left(n\right)} & 0 & 0 \\
0 & \Phi^{\left(n\right)} & 0 \\
0 & 0 & \Phi^{\left(n\right)}%
\end{array}%
\right)  \notag
\\
&+\frac{1}{2}\left(%
\begin{array}{ccc}
\left(\left(q_{n}^{\Psi}\right)^{2}-1\right)\Psi_{D}^{\left(n\right)} & 0 &
\Psi_{B}^{\left(n\right)} \\
0 & \left(\left(q_{n}^{\Psi}\right)^{2}-1\right)\Psi_{D}^{\left(n\right)} &
\Psi_{C}^{\left(n\right)} \\
\Psi_{B}^{\left(n\right)} & \Psi_{C}^{\left(n\right)} & 2\Psi_{D}^{\left(n%
\right)}%
\end{array}%
\right).  \label{eq:h_ij-decomposed}
\end{align}
Thus, an idealized spherical shell of multiple test masses set in motion by
a passing gravitational wave is able to detect six modes: one
scalar mode $\left(\Phi\right)$, two transverse modes $\left(h_{+}+%
\Psi_{+},h_{\times}+\Psi_{\times}\right)$, and three longitudinal modes $%
\left(\Psi_{B},\Psi_{C},\Psi_{D}\right)$.

\subsection{Total power radiated\label{subsec:Total-power}}

Herein we shall obtain the total power radiated $P=dE/dt$ in terms of the
modes described in Section \ref{subsec:general-direction}.

The quantities to be calculated appearing in Eq. (\ref{eq:energy-loss}) are
invariants under Lorentz transformations. Accordingly, a rotation can be
performed in such a way that the $z$-direction corresponds to the observer's
line of sight. In this situation, we fix the gauge for $\tilde{h}%
_{\mu\nu}^{\left(n\right)}$ and $\Psi_{\mu\nu}^{\left(n\right)}$ so that
\begin{eqnarray}
\partial^{\mu}\tilde{h}_{\mu\nu}^{\left(n\right)}=\partial^{\mu}\Psi_{\mu%
\nu}^{\left(n\right)}=0, \\
\text{ \ }\tilde{h}^{\left(n\right)}=\Psi^{\left(n%
\right)}=0\text{ \ and \ }\tilde{h}_{0\nu}^{\left(n\right)}=0,
\label{eq:h-tilde-Psi-gauge-fix}
\end{eqnarray}

thus obtaining
\begin{equation}
\tilde{h}_{\mu\nu}^{\left(n\right)}=\left(%
\begin{array}{cccc}
0 & 0 & 0 & 0 \\
0 & h_{+}^{\left(n\right)} & h_{\times}^{\left(n\right)} & 0 \\
0 & h_{\times}^{\left(n\right)} & -h_{+}^{\left(n\right)} & 0 \\
0 & 0 & 0 & 0%
\end{array}%
\right)  \label{eq:h-tilde-matrix}
\end{equation}
and
\begin{widetext}
\begin{equation}
\Psi_{\mu\nu}^{\left(n\right)}=\left(%
\begin{array}{cccc}
\left(q_{n}^{\Psi}\right)^{2}\Psi_{D}^{\left(n\right)} & q_{n}^{\Psi}%
\Psi_{B}^{\left(n\right)} & q_{n}^{\Psi}\Psi_{C}^{\left(n\right)} &
q_{n}^{\Psi}\Psi_{D}^{\left(n\right)} \\
q_{n}^{\Psi}\Psi_{B}^{\left(n\right)} & \frac{\left(\left(q_{n}^{\Psi}%
\right)^{2}-1\right)}{2}\Psi_{D}^{\left(n\right)}+\Psi_{+}^{\left(n\right)}
& \Psi_{\times}^{\left(n\right)} & \Psi_{B}^{\left(n\right)} \\
q_{n}^{\Psi}\Psi_{C}^{\left(n\right)} & \Psi_{\times}^{\left(n\right)} &
\frac{\left(\left(q_{n}^{\Psi}\right)^{2}-1\right)}{2}\Psi_{D}^{\left(n%
\right)}-\Psi_{+}^{\left(n\right)} & \Psi_{C}^{\left(n\right)} \\
q_{n}^{\Psi}\Psi_{D}^{\left(n\right)} & \Psi_{B}^{\left(n\right)} &
\Psi_{C}^{\left(n\right)} & \Psi_{D}^{\left(n\right)}%
\end{array}%
\right).  \label{eq:Psi-matriz}
\end{equation}
\end{widetext}
The massless spin-2 mode is then subjected to
\begin{equation}
\partial_{t}\tilde{h}_{\left(n\right)}^{\alpha\beta}\partial_{t}\tilde{h}
_{\alpha\beta}^{\left(n\right)}=2\left[\left(\dot{h}_{+}^{\left(n\right)}
\right)^{2}+\left(\dot{h}_{\times}^{\left(n\right)}\right)^{2}\right],
\label{eq:h-dot-squared}
\end{equation}
while the massive spin-2 mode satisfies

\begin{align}
\dot{\Psi}_{\left( n\right) }^{\alpha \beta }\dot{\Psi}_{\alpha \beta
}^{\left( n\right) } &=2\left( \dot{\Psi}_{+}^{\left( n\right) }\right)
^{2}+2\left( \dot{\Psi}_{\times }^{\left( n\right) }\right) ^{2}+2\left( 
\dot{\Psi}_{B}^{\left( n\right) }\right) ^{2}  \notag \\
&-2\left( q_{n}^{\Psi }\right) ^{2}\left( \dot{\Psi}_{B}^{\left( n\right)
}\right) ^{2} 
 -2\left( q_{n}^{\Psi
}\right) ^{2}\left( \dot{\Psi}_{C}^{\left( n\right) }\right) ^{2} \notag \\
&  +2\left( \dot{\Psi}_{C}^{\left( n\right) }\right) ^{2} 
+\frac{3}{2}\left( 1-\left( q_{n}^{\Psi }\right) ^{2}\right) ^{2}\left( 
\dot{\Psi}_{D}^{\left( n\right) }\right) ^{2} . \label{eq:Psi-dot-squared}
\end{align}
Eqs. (\ref{eq:h-dot-squared}) and (\ref{eq:Psi-dot-squared}) can be inserted
into Eq. (\ref{eq:energy-loss}). This yields the expression of the total power radiated in terms of the propagation modes:
\begin{align}
P& =\frac{c^{3}r^{2}}{16\pi G}\sum_{n}\int d\Omega \left\{ \left\langle
\left( \dot{h}_{+}^{\left( n\right) }\right) ^{2}\right\rangle +\left\langle
\left( \dot{h}_{\times }^{\left( n\right) }\right) ^{2}\right\rangle \right. 
\notag \\
& -q_{n}^{\Psi }\left[ \left\langle \left( \dot{\Psi}_{+}^{\left( n\right)
}\right) ^{2}\right\rangle +\left\langle \left( \dot{\Psi}_{\times }^{\left(
n\right) }\right) ^{2}\right\rangle \right] \Theta _{n}^{\Psi }  \notag \\
& +3q_{n}^{\Phi }\left\langle \left( \dot{\Phi}^{\left( n\right) }\right)
^{2}\right\rangle \Theta _{n}^{\Phi }  \notag \\
& -q_{n}^{\Psi }\left( 1-\left( q_{n}^{\Psi }\right) ^{2}\right) \left[
\left\langle \left( \dot{\Psi}_{B}^{\left( n\right) }\right)
^{2}\right\rangle +\left\langle \left( \dot{\Psi}_{C}^{\left( n\right)
}\right) ^{2}\right\rangle \right] \Theta _{n}^{\Psi }
\notag \\
& \left. -\frac{3}{4}q_{n}^{\Psi }\left( 1-\left( q_{n}^{\Psi }\right)
^{2}\right) ^{2}\left\langle \left( \dot{\Psi}_{D}^{\left( n\right) }\right)
^{2}\right\rangle \Theta _{n}^{\Psi }\right\} . \label{eq:P(h-dot,Psi-dot)} 
\end{align}%
Due to the presence of the massive spin-2 modes, the above quantity is not
positive definite. This feature is potentially problematic because a
radiated power in the form of gravitational waves must be non-negative. In
order to better characterize this possible pathology, we shall write $P$ in
the quadrupole approximation.

The integration with respect to the angular variables in $d\Omega$ in Eq. (%
\ref{eq:P(h-dot,Psi-dot)}) instructs us to account for the various modes
propagating in an arbitrary direction. By inserting Eqs. (\ref{eq:h_plus}), (%
\ref{eq:h_cross}), (\ref{eq:Psi_plus}), (\ref{eq:Psi_cross}), (\ref{eq:Psi_B}%
), (\ref{eq:Psi_C}), (\ref{eq:Psi_D}), and (\ref{eq:Phi}) into (\ref%
{eq:P(h-dot,Psi-dot)}), we obtain, after a long calculation, the following:
\begin{align}
& P =\frac{G}{5c^{5}}\sum_{n}\left\{ \left\langle \dddot{Q}_{ij}^{\left(
n\right) }\dddot{Q}_{ij}^{\left( n\right) }\right\rangle \right.  \notag \\
& -q_{n}^{\Psi }\left\langle \left( 1-\hat{O}_{\Psi }\right) \dddot{Q}%
_{ij}^{\left( n\right) }\left( 1-\hat{O}_{\Psi }\right) \dddot{Q}%
_{ij}^{\left( n\right) }\right\rangle \Theta _{n}^{\Psi }  \notag \\
& +\frac{1}{18}q_{n}^{\Phi }\left\langle \left( 1-\hat{O}_{\Phi }\right) 
\dddot{K}_{ij}^{\left( n\right) }\left( 1-\hat{O}_{\Phi }\right) \dddot{K}%
_{ij}^{\left( n\right) }\right\rangle \Theta _{n}^{\Phi }  \notag \\
& -q_{n}^{\Psi }\left[ \left( \frac{m_{\Psi }c}{\omega _{n}}\right)
^{2}\left\langle \left( 1-\hat{O}_{\Psi }\right) \dddot{Q}_{ij}^{\left(
n\right) }\left( 1-\hat{O}_{\Psi }\right) \dddot{Q}_{ij}^{\left( n\right)
}\right\rangle \right.   \notag \\
& \left. \left. +\frac{1}{2}\left( \frac{m_{\Psi }c}{\omega _{n}}\right)
^{4}\left\langle \left( 1-\hat{O}_{\Psi }\right) \dddot{K}_{ij}^{\left(
n\right) }\left( 1-\hat{O}_{\Psi }\right) \dddot{K}_{ij}^{\left( n\right)
}\right\rangle \right] \Theta _{n}^{\Psi }\right\}, \label{eq:P(Q,K,O-hat)}
\end{align}%
where
\begin{align}
Q_{ij} & \equiv M_{ij}-\frac{1}{3}\delta_{ij}M_{kk},  \label{eq:Qij} \\
K_{ij} & \equiv M_{ij}-\frac{1}{3}\left[1+\sqrt{\frac{5}{2}}\right]%
\delta_{ij}M_{kk}.  \label{eq:Kij}
\end{align}
We track the contribution of each propagation mode to the various terms in
Eq. (\ref{eq:P(Q,K,O-hat)}) and establish the following association
\begin{equation}
\left\langle \dddot{Q}_{ij}^{\left(n\right)}\dddot{Q}_{ij}^{\left(n\right)}
\right\rangle \rightarrow h_{+,\times},  \notag
\end{equation}
\begin{equation}
q_{n}^{\Psi} \left\langle \left(1-\hat{O}_{\Psi}\right)\dddot{Q}%
_{ij}^{\left(n\right)}\left(1-\hat{O}_{\Psi}\right)\dddot{Q}%
_{ij}^{\left(n\right)}\right\rangle \rightarrow\Psi_{+,\times},
\notag
\end{equation}
\begin{equation}
q_{n}^{\Psi}\left(\frac{m_{\Psi}c}{\omega_{n}}\right)^{2} \left\langle
\left(1-\hat{O}_{\Psi}\right)\dddot{Q}_{ij}^{\left(n\right)}\left(1-\hat{O}_{\Psi}\right)\dddot{Q}_{ij}^{\left(n\right)}\right\rangle  \rightarrow\Psi_{B,C},  \notag
\end{equation}
\begin{equation}
 q_{n}^{\Psi}\left(\frac{m_{\Psi}c}{\omega_{n}}\right)^{4} \left\langle
\left(1-\hat{O}_{\Psi}\right)\dddot{K}_{ij}^{\left(n\right)}\left(1-\hat{O}_{\Psi}\right)\dddot{K}_{ij}^{\left(n\right)}\right\rangle 
\rightarrow\Psi_{D},  \notag
\end{equation}
\begin{eqnarray}
q_{n}^{\Phi}\left\langle \left(1-\hat{O}_{\Phi}\right)\dddot{K}%
_{ij}^{\left(n\right)}\left(1-\hat{O}_{\Phi}\right)\dddot{K}%
_{ij}^{\left(n\right)}\right\rangle & \rightarrow\Phi.
\notag
\end{eqnarray}
Here, the symbol ``$\rightarrow$'' has the meaning ``is related to'' or ``comes from''.

Eq. (\ref{eq:P(Q,K,O-hat)}) may be simplified if we consider $P$ in the
radiation zone, wherein
\begin{align}
\dddot{Q}_{ij}^{\left(n\right)} & \sim\frac{1}{r}\dot{q}_{ij}^{\left(n%
\right)}\left(\omega_{n}\left(t-r/c\right)\right),  \label{eq:Qij-rad} \\
\left(1-\hat{O}_{\Psi}\right)\dddot{Q}_{ij}^{\left(n\right)} & \sim\frac{1}{r%
}\dot{q}_{ij}^{\left(n\right)}\left(\omega_{n}\left(t-r/v_{\Psi}^{\left(n%
\right)}\right)\right),  \label{eq:O-hat-Psi-Qij-rad}
\end{align}
and
\begin{align}
\left(1-\hat{O}_{\Phi}\right)\dddot{K}_{ij}^{\left(n\right)} & \sim\frac{1}{r%
}\dot{k}_{ij}^{\left(n\right)}\left(\omega_{n}\left(t-r/v_{\Phi}^{\left(n%
\right)}\right)\right),  \label{eq:O-hat-Phi-Kij-rad} \\
\left(1-\hat{O}_{\Psi}\right)\dddot{K}_{ij}^{\left(n\right)} & \sim\frac{1}{r%
}\dot{k}_{ij}^{\left(n\right)}\left(\omega_{n}\left(t-r/v_{\Psi}^{\left(n%
\right)}\right)\right).  \label{eq:O-hat-Psi-Kij-rad}
\end{align}

Within the average sign $\left\langle ...\right\rangle $, the quadratic
combinations of the pairs ((\ref{eq:Qij-rad}),(\ref{eq:O-hat-Psi-Qij-rad}))
and ((\ref{eq:O-hat-Phi-Kij-rad}),(\ref{eq:O-hat-Psi-Kij-rad})) give the
same result independently of the physical system being described. Therefore,
Eq. (\ref{eq:P(Q,K,O-hat)}) can be cast into the form
\begin{align}
P& =\frac{G}{5c^{5}}\sum_{n}\left\{ \left( 1-q_{n}^{\Psi }\Theta _{n}^{\Psi
}\right) \left\langle \dddot{Q}_{ij}^{\left( n\right) }\dddot{Q}%
_{ij}^{\left( n\right) }\right\rangle \right.   \notag \\
& +\frac{1}{18}q_{n}^{\Phi }\left\langle \dddot{K}_{ij}^{\left( n\right) }%
\dddot{K}_{ij}^{\left( n\right) }\right\rangle \Theta _{n}^{\Phi }  \notag \\
& -q_{n}^{\Psi }\left( \frac{m_{\Psi }c}{\omega _{n}}\right)
^{2}\left\langle \dddot{Q}_{ij}^{\left( n\right) }\dddot{Q}_{ij}^{\left(
n\right) }\right\rangle \Theta _{n}^{\Psi }  \notag \\
& \left. -\frac{1}{2}q_{n}^{\Psi }\left( \frac{m_{\Psi }c}{\omega _{n}}%
\right) ^{4}\left\langle \dddot{K}_{ij}^{\left( n\right) }\dddot{K}%
_{ij}^{\left( n\right) }\right\rangle \Theta _{n}^{\Psi }\right\} .
\label{eq:P(Q,K)}
\end{align}%
The sign of $P$ is the concerning point regarding Eq. (\ref{eq:P(Q,K)}). Its first two lines are positive definite, since $0\leq q_{n}^{X}\leq1$ (with $X=\Psi$,%
$\Phi$). In fact, the modes $\left\{ h_{+},\Psi_{+}\right\} $ and $\left\{
h_{\times},\Psi_{\times}\right\} $ combine in a pattern of destructive
interference. The same is not true for the terms in the last two lines of Eq. (%
\ref{eq:P(Q,K)}) with relation to the terms involving the longitudinal modes
$\Psi_{B}$, $\Psi_{C}$, and $\Psi_{D}$. The contribution of the latter modes
for $P$ will always be negative.

We will show that the non-positivity of $P$ creates inconsistencies in the
description of rather simple physical systems.

\subsection{Binary point-mass system in circular orbit\label{subsec:BBH}}

One of the simplest (and most interesting) physical systems to be studied is
the set of two point masses $m_{1}$ and $m_{2}$ in mutual circular orbit
with radius $R$ and angular frequency $\omega_{s}$. The trajectories of both
particles define a plane of motion, which may be called the $xy$-plane.
Accordingly, the non-null mass moments are $M_{11}=-M_{22}$ and $M_{12}$,
only \cite{Maggiore2007}. Under the assumption of a non-relativistic motion,
it can be shown that the emitted gravitational wave is monochromatic of
frequency $\omega=2\omega_{s}$; it is also true that \cite{Alves2022}:
\begin{equation}
\left\langle \dddot{Q}_{ij}\dddot{Q}_{ij}\right\rangle =\left\langle \dddot{K%
}_{ij}\dddot{K}_{ij}\right\rangle =32\mu^{2}R^{4}\omega_{s}^{6},
\label{eq:averaged-Qij2-BBH}
\end{equation}
where $\mu=\frac{m_{1}m_{2}}{m_{1}+m_{2}}$ is the reduced mass.

For simplicity reasons, let us consider that only the tensor modes are
emitted by the binary system while the scalar mode is in the damping regime.
Consequently,
\begin{equation}
\Theta^{\Psi}=1\quad\text{and}\quad\Theta^{\Phi}=0\text{,}
\label{eq:Theta-tensor-emission}
\end{equation}
and the total power radiated (\ref{eq:P(Q,K)}) reads:
\begin{align}
P &= P_{\text{GR}}\left[ 1-\sqrt{1-\left( \frac{m_{\Psi }c}{2\omega _{s}} \right) ^{2}}\left( 1+\left( \frac{m_{\Psi }c}{2\omega _{s}}\right)
^{2}\right. \right.  \notag  \\
& \qquad \quad \left. \left. +\frac{1}{2}\left( \frac{m_{\Psi }c}{2\omega _{s}}\right)
^{4}\right) \right] ,  \label{eq:P-BBH}
\end{align}%
with
\begin{equation}
P_{\text{GR}}=\frac{32G\mu^{2}R^{4}\omega_{s}^{6}}{5c^{5}},  \label{eq:P_GR}
\end{equation}
the ordinary expression from general relativity.

\begin{figure}[h!]
\begin{centering}
\includegraphics[scale=0.65]{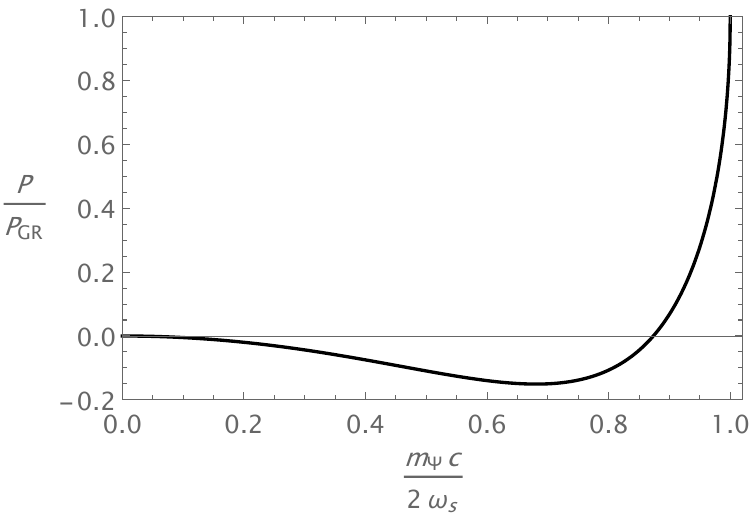}
\par\end{centering}
\caption{Binary system radiated power $P$ in units of $P_{\text{GR}}$ as a
function of $\frac{m_{\Psi}c}{2\protect\omega_{s}}$. The region within the
interval $0<\frac{m_{\Psi}c}{2\protect\omega_{s}}<0.87$ displays $P<0$.}
\label{fig:P-BBH}
\end{figure}

The plot of Eq. (\ref{eq:P-BBH})---see Fig. \ref{fig:P-BBH}---clearly
displays the pathological region where the power is negative-valued. Our
next necessary step is to verify if this problematic region is accessed
dynamically.

The dynamics of the binary system are determined by the energy balance equation
\begin{equation}
P=-\frac{dE_{\text{orbit}}}{dt},  \label{eq:balance-eq}
\end{equation}
where $E_{\text{orbit}}$ is the Newtonian orbital energy\footnote{%
The non-relativistic binary black hole systems are described by Newtonian dynamics in quadratic gravity \cite{Lu2015}.}
\begin{equation}
E_{\text{orbit}}=-\frac{G\mu m}{2R}  \label{eq:E_orb}
\end{equation}
with $m=m_{1}+m_{2}$.

We differentiate Eq. (\ref{eq:E_orb}) with respect to time and make use of
the third Kepler's law for circular orbits, $\omega_{s}^{2}=Gm/R^{3}$ , thus
obtaining
\begin{equation}
\frac{dE_{\text{orbit}}}{dt}=-\frac{\left(GM_{c}\right)^{5/3}}{3G}\dot{\omega%
}_{s}\omega_{s}^{-1/3}.  \label{eq:E_orb-dot}
\end{equation}
Here $M_{c}=m^{\frac{2}{5}}\mu^{\frac{3}{5}}$ is the chirp mass.

Eqs. (\ref{eq:P-BBH}) and (\ref{eq:E_orb}) are substituted into Eq. (\ref%
{eq:balance-eq}), yielding:
\begin{align}
\dot{\omega}_{s} &= \omega _{\text{GR}}\left\{ 1-\sqrt{1-\left( \frac{m_{\Psi}c}{2\omega _{s}}\right) ^{2}}\left[ 1+\left( \frac{m_{\Psi }c}{2\omega _{s}}%
\right) ^{2}\right. \right.   \notag \\
& \qquad \quad \left. \left. +\frac{1}{2}\left( \frac{m_{\Psi }c}{2\omega _{s}}\right)
^{4}\right] \right\} , \label{eq:omega_s-dot}
\end{align}%
with
\begin{equation}
\omega _{GR}=\frac{96}{5}\left( \frac{GM_{c}}{c^{3}}\right) ^{5/3}\omega
_{s}^{11/3}.
\end{equation}%
The integration constant present in the solution of this first-order
ordinary differential equation is set by the initial condition $%
\omega_{s}\left(0\right)\equiv\omega_{0}=\frac{m_{\Psi}c}{2}$. This
condition says that the system starts emitting the massive spin-2 modes at $%
t=0$. The solution to Eq. (\ref{eq:omega_s-dot}) is plotted in Fig. \ref%
{fig:omega_s-BBH}.
\begin{figure}[h!]
\begin{centering}
\includegraphics[scale=0.65]{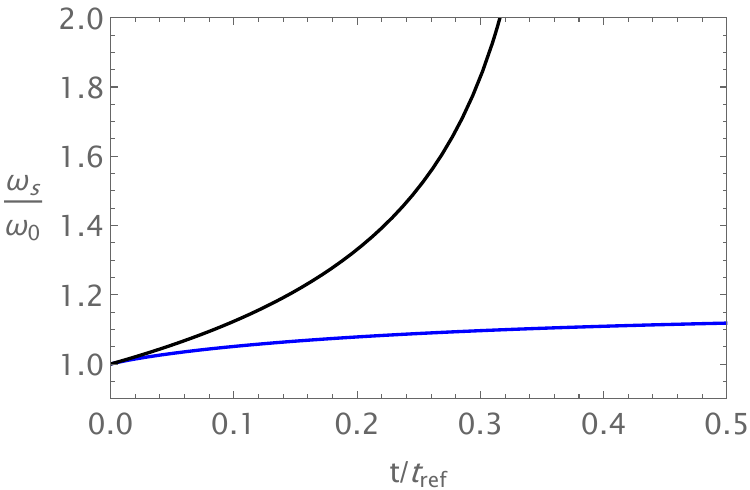}
\par\end{centering}
\caption{Orbital angular frequency as a function of time for the complete
spin-2 case---blue curve---and the massless spin-2 mode of GR---black curve.
The reference time is given by $t_{\text{ref}}=\frac{5}{3}2^{\frac{1}{3}%
}\left(\frac{GM_{c}}{c^{3}}\right)^{-\frac{5}{3}}\left(2m_{\Psi}c\right)^{-%
\frac{8}{3}}$.}
\label{fig:omega_s-BBH}
\end{figure}

{}

The main distinction between the curves displayed in Fig. \ref%
{fig:omega_s-BBH} is the following: the angular frequency as a function of
time in GR is the black curve; it diverges in the finite time $\frac{t}{t_{%
\text{ref}}}=\frac{3}{8}$. The blue curve gives $\omega_{s}\left(t\right)$
for the complete spin-2 mode; it approaches the constant value $%
\omega_{s}^{\max}\simeq1.145\omega_{0}$ as $t\rightarrow\infty$. The finite
value of the angular frequency for the complete case occurs in the limit of
the square bracket of Eq. (\ref{eq:omega_s-dot}) approaching zero, i.e. for
the power $P=0$. Therefore, in this particular case, the total power
radiated will never assume negative values.

Na\"{\i}vely, the above result could be interpreted in this way: initially,
the system emits only $h$-type modes thus exhibiting a dynamics identical to
GR's; when the frequency $\omega_{s}$ reaches the value $\omega_{0}$, the $%
\Psi$-type modes are activated and $h$- and $\Psi$-type modes combine in an
interference pattern with a progressively destructive effect; $\omega_{s}$
approaches the value $\omega_{s}^{\text{max}}$ as $t\rightarrow\infty$, i.e.
the complete destructive interference takes place in the asymptotic limit of
time.

However, a careful consideration of the behavior of the various $h$- and $%
\Psi$-modes in the section \ref{subsec:general-direction} shows that the
interpretation in the paragraph above is not correct. Take, for example, the
limiting case where $\omega_{s}=\omega_{s}^{\max}$ and $P=0$. Therein, the
destructive interference should be complete in every direction. Yet, for the
direction $\theta=\phi=0$, the aforementioned modes read:

\begin{align}
h_{+}\left( \mathbf{\hat{z}}\right) & =\frac{\kappa }{4\pi r}\ddot{M}_{11} \notag \\
& =A_{\max }\cos \left[ 2\omega _{s}^{\max }\left( t-\frac{r}{c}\right) %
\right] ,  \label{eq:h_plus_lim_direct} \\
h_{\times }\left( \mathbf{\hat{z}}\right) & =\frac{\kappa }{4\pi r}\ddot{M}%
_{12}  \notag \\
& =A_{\max }\sin \left[ 2\omega _{s}^{\max }\left( t-\frac{r}{c}\right) %
\right] ,  \label{eq:h_cross_lim_direct}
\end{align}%
and 
\begin{align}
\Psi _{+}\left( \mathbf{\hat{z}}\right) & =\frac{\kappa }{4\pi r}\left[ 1-%
\hat{O}_{\Psi }\right] \ddot{M}_{11} \notag  \\
& =-A_{\max }\cos \left[ 2\omega _{s}^{\max }\left( t-\left( \frac{r}{c}%
\right) q_{\max }^{\Psi }\right) \right] ,   \label{eq:Psi_plus_lim_direct} \\
\Psi _{\times }\left( \mathbf{\hat{z}}\right) & =\frac{\kappa }{4\pi r}\left[
1-\hat{O}_{\Psi }\right] \ddot{M}_{12}  \notag \\
& =-A_{\max }\sin \left[ 2\omega _{s}^{\max }\left( t-\left( \frac{r}{c}%
\right) q_{\max }^{\Psi }\right) \right] , \label{eq:Psi_cross_lim_direct} \\
\Psi _{B}\left( \mathbf{\hat{z}}\right) & =\Psi _{C}\left( \mathbf{\hat{z}}%
\right) =\Psi _{D}\left( \mathbf{\hat{z}}\right) =0,
\label{Psi_long_lim_direct}
\end{align}%
where 
\begin{align}
A_{\max }& =\frac{4c}{r}\left( \frac{GM_{c}}{c^{3}}\right) ^{\frac{5}{3}%
}\left( \omega _{s}^{\max }\right) ^{\frac{2}{3}},  \label{eq:A_max} \\
q_{\max }^{\Psi }& =\sqrt{1-\left( \frac{m_{\Psi }c}{2\omega _{s}^{\max }}%
\right) ^{2}}\simeq 0.49.  \label{eq:q_max}
\end{align}%

The results in Eqs. (\ref{eq:h_plus_lim_direct}), (\ref%
{eq:h_cross_lim_direct}), (\ref{eq:Psi_plus_lim_direct}), and (\ref%
{eq:Psi_cross_lim_direct}) were explicitly obtained in Ref. \cite{Alves2022}.

The destructive interference is not complete because $q_{\max}^{\Psi}\neq1$.
For instance, for $t\gg t_{\text{ref}}$:
\begin{align}
h_{+}\left(\mathbf{\hat{z}}\right)+\Psi_{+}\left(\mathbf{\hat{z}}\right) &
=B\sin\left[2\omega_{s}^{\max}\left(t-\frac{r}{2c}\left(1+q_{\max}^{\Psi}%
\right)\right)\right],  \notag \\
h_{\times}\left(\mathbf{\hat{z}}\right)+\Psi_{\times}\left(\mathbf{\hat{z}}%
\right) & =-B\cos\left[2\omega_{s}^{\max}\left(t-\frac{r}{2c}%
\left(1+q_{\max}^{\Psi}\right)\right)\right],  \notag
\end{align}
with
\begin{equation}
B=2A_{\max}\sin\left[\left(\frac{r}{c}\right)\omega_{s}^{\max}\left(1-q_{%
\max}^{\Psi}\right)\right].  \label{eq:B}
\end{equation}
This is a situation where the total power radiated is null; nevertheless, an
observer with line-of-sight along the $\mathbf{\hat{z}}$-direction would
detect gravitational waves of amplitude $B$.

The previous analysis clearly shows a physical inconsistency in the modified
gravitational model even with $P$ being always positive valued. This inconsistency also arises in the calculation of the radiated angular momentum given by Eq. (\ref{eq:angular-momentum-loss}). Notably, the sign difference between the terms in Eqs. (\ref{eq:j-i-h}) and (\ref{eq:j-i-Psi}) highlights the presence of Ostrogradsky instability in $j_{\Psi }^{i\left( n\right) }$. Furthermore, the term $\dot{\Psi}_{\left( n\right) }^{\alpha \beta
}x_{k}\partial _{l}\Psi _{\alpha \beta }^{\left( n\right) }$ in Eq. (\ref{eq:j-i-Psi}) contains both transverse and longitudinal modes. Consequently, even when considering the term $\dot{h}_{ab}^{\left( n\right) }x_{k}\partial_{l}h_{ab}^{\left( n\right) }$ in Eq. (\ref{eq:j-i-h}), it is not possible to combine $h_{ab}^{\left( n\right) }$ and $\Psi _{\alpha \beta }^{\left( n\right) }$ in a way that ensures the positivity of Eq. (\ref{eq:angular-momentum-loss}).

The next section makes it clear that the issue described previously is related to the presence of longitudinal modes in the quadrupole approximation.

\subsection{Transverse-traceless spin$-2$ massive modes\label%
{subsec:TTmassivespin2} \ }

The problem revealed in the previous sections is incepted by the term $%
C_{\mu\nu\alpha\beta}C^{\mu\nu\alpha\beta}$ in the Lagrangian of the
quadratic gravity. In fact, it is long known in the literature \cite%
{Stelle1977} that such a term presents an Ostrogradsky instability leading
to non-positive definite energy. In the context of gravitational waves, this
translates to negative-valued radiated power, a clear physical inconsistency.

One way to remedy this issue is to argue that the massive spin-2 mode will
manifest itself exclusively in the damping regime thus avoiding pathological
propagating modes. This is equivalent to state that $m_{\Psi}$ very large so
that $\frac{c}{\omega}>m_{\Psi}^{-1}$ in the classical length scales. In
this way, the term $C_{\mu\nu\alpha\beta}C^{\mu\nu\alpha\beta}$ will be
important in the quantum domain while being completely negligible in the
classical context of gravitational waves.\footnote{%
In the quantum realm, Ostrogradsky instability translates to the existence
of ghost-like fields. Even though ghost fields are problematic \cite%
{Sbisa2015}, there are approaches in the literature dedicated to dealing
consistently with this type of field \cite%
{Modesto2016,Donoghue2019,Salvio2019,Salvio2021,Anselmi2017,Anselmi2018}.}
Another way to deal with this problem is to impose that the
transverse-traceless massive spin-2 modes are the only modes present 
in the
theory. In other words, there should be
\begin{equation}
\Psi_{ij}\rightarrow\Lambda_{ij,kl}\Psi_{kl}=\Psi_{ij}^{TT}.
\label{eq:Psi-TT}
\end{equation}
Under this assumption, the longitudinal modes $\Psi_{B}$, $\Psi_{C}$, 
and $
\Psi_{D}$ are disregarded and a consistent scenario of interference patterns involving $\left\{ h_{+},\Psi_{+}\right\} $ and $\left\{
h_{\times},\Psi_{\times}\right\} $ naturally takes place. Additionally, all
the terms in the second lines of Eqs. (\ref{eq:P(h-dot,Psi-dot)}), (\ref%
{eq:P(Q,K,O-hat)}), and (\ref{eq:P(Q,K)}) vanish, thus leading to a
positive-definite power.

Under the prescription (\ref{eq:Psi-TT}), the energy-loss equation (\ref%
{eq:energy-loss}) and the angular-momentum-loss equation (\ref%
{eq:angular-momentum-loss}) reduce to:
\begin{align}
\frac{dE}{dt} &= \frac{c^{3}r^{2}}{32\pi G}\sum_{n}\int d\Omega \left[
\left\langle \dot{h}_{ab}^{TT}\dot{h}_{ab}^{TT}\right\rangle _{\left(
n\right) }\right.  \notag \\
& -q_{n}^{\Psi }\Theta _{n}^{\Psi }\left\langle \dot{\Psi}_{ab}^{TT}\dot{\Psi%
}_{ab}^{TT}\right\rangle _{\left( n\right) }+\left. 6q_{n}^{\Phi }\Theta
_{n}^{\Phi }\left\langle \dot{\Phi}^{2}\right\rangle _{\left( n\right) }%
\right] ,  \label{eq:energy-loss-TT}
\end{align}%
and 
\begin{align}
\frac{dJ^{i}}{dt}& =\frac{c^{3}r^{2}}{32\pi G}\sum_{n}\epsilon ^{ikl}\int
d\Omega \left\{ -\left\langle \dot{h}_{ab}^{TT}x_{k}\partial
_{l}h_{ab}^{TT}\right\rangle _{\left( n\right) }\right.  \notag \\
& +2\left\langle h_{ak}^{TT}\dot{h}_{al}^{TT}\right\rangle _{\left( n\right)
}+q_{n}^{\Psi }\Theta _{n}^{\Psi }\left\langle \dot{\Psi}_{ab}^{TT}x_{k}%
\partial _{l}\Psi _{ab}^{TT}\right\rangle _{\left( n\right) }  \notag \\
& \left. -2q_{n}^{\Psi }\Theta _{n}^{\Psi }\left\langle \Psi _{ak}^{TT}\dot{%
\Psi}_{al}^{TT}\right\rangle _{\left( n\right) }-6q_{n}^{\Phi }\Theta
_{n}^{\Phi }\left\langle \dot{\Phi}x_{k}\partial _{l}\Phi \right\rangle
_{\left( n\right) }\right\}. \label{eq:angular-momentum-loss-TT} 
\end{align}%
It should be emphasized that, although the $h$-mode and the $\Psi$-mode are
both transverse-traceless fields, the origins of these restrictions are
distinct: for massless modes the TT restriction arises from the imposition
of a gauge that selects only the physical degrees of freedom; for massive
modes the TT condition is necessary to eliminate the instability of the
propagating modes in the quadrupole approximation.

The description of a binary system of point-like masses in circular orbits
under the prescription (\ref{eq:Psi-TT}) was studied by the authors of Ref. \cite{Alves2022} in
detail. In that reference, the physical consistency of the system was shown
in the context where the gravitational waves carry energy away, only. In the
next section, we study another example where the gravitational waves
emission carries away both energy and angular momentum.

\section{Rotating ellipsoid with free precession\label%
{sec:Rotating-ellipsoid}}

This section is devoted to the study of an astrophysical object that may be
described as an ellipsoid rotating with free precession. In most
astronomical objects, the rotation axis typically does not align with a
principal axis. As a result, the motion of the rigid body involves a
combination of rotation around a principal axis and precession. As is well
explored in \cite{Maggiore2007}, this precession motion introduces
qualitatively new characteristics into the emitted gravitational waves. As
we will see, these waves reduce both the energy and angular momentum of the
body.

In order to calculate these gravitational waves, we introduce a fixed
reference frame $\left( x_{1},x_{2},x_{3}\right) $. In this inertial frame,
the angular momentum $\vec{J}$ of the rigid body is conserved, and we align
the $x_{3}$ axis with $\vec{J}$. Additionally, we introduce the body's
reference frame, which is attached to the body with coordinates $\left(
x_{1}^{\prime} ,x_{2}^{\prime},x_{3}^{\prime}\right) $, whose axes are
aligned with the principal axes of the body. The relationship between the
two reference frames, described by the Euler angles $\left(
\alpha,\beta,\gamma\right) $, connects the coordinates of the frames through
the equation $x_{i}^{\prime}=\mathcal{R}_{ij}x_{j}$, where $\mathcal{R}_{ij}$
is the rotation matrix given by
\begin{widetext}
\begin{equation}
\mathcal{R}_{ij}=\left(
\begin{array}{ccc}
\cos\gamma & \sin\gamma & 0 \\
-\sin\gamma & \cos\gamma & 0 \\
0 & 0 & 1%
\end{array}
\right) \left(
\begin{array}{ccc}
1 & 0 & 0 \\
0 & \cos\alpha & \sin\alpha \\
0 & -\sin\alpha & \cos\alpha%
\end{array}
\right) \left(
\begin{array}{ccc}
\cos\beta & \sin\beta & 0 \\
-\sin\beta & \cos\beta & 0 \\
0 & 0 & 1%
\end{array}
\right) .  \label{rotationMatrix}
\end{equation}
\end{widetext}

 The motion of the rigid body is specified once we know how $\alpha$%
, $\beta$ and $\gamma$ evolve with time.

As is well known, gravitational waves are produced by the body's
time-varying quadrupole moment. The components of the inertia tensor $\
I_{ij}$ and the reduced form of the quadrupole moment $Q_{ij}$ are defined
as follows:%
\begin{equation}
I_{ij}=\int\left( r^{2}\delta_{ij}-x_{i}x_{j}\right) \rho dV,
\label{inertial_mon}
\end{equation}%
\begin{equation}
Q_{ij}=M_{ij}-\frac{1}{3}\delta_{ij}M_{k}^{k}=\int\left(
x_{i}x_{j}-r^{2}\delta_{ij}\right) \rho dV,  \label{quadrupole_mon}
\end{equation}
where $\rho$ is the mass volume density of the body. Note that definitions (%
\ref{inertial_mon}) and (\ref{quadrupole_mon}) are equivalent, apart from a
sign. This leads to the relation $M_{ij}=-I_{ij}+c_{ij}$, where $c_{ij}$ are
constants. These constants are irrelevant when we consider the time
derivatives that will characterize the gravitational wave.

Consider an axisymmetric body whose longitudinal axis $x_{3}^{\prime}$ forms
an angle $\alpha$ with the axis of angular momentum $x_{3}$. This angle $%
\alpha $ is referred to as the wobble angle and the gravitational wave
emission associated with this motion is known as wobble radiation \cite%
{Maggiore2007}. When the rotation axis does not align with the principal
axis of the body, the angular momentum rotates around the principal axis, leading to precession. According to \cite{Marion2014}, by solving the force-free Euler equations, it is evident that free precession occurs as the principal axis rotates around the angular momentum axis with a frequency of

\begin{equation}
\Omega \equiv \dot{\beta}=\frac{J}{I_{1}},  \label{omega_beta}
\end{equation}%
where the angular momentum is $\vec{J}=\left( 0,0,J\right) $. In Section \ref%
{subsec:general-direction}, we derived the propagation modes for
gravitational waves in a generic direction, considering all polarizations of
the fields $h_{\mu \nu }$, $\Psi _{\mu \nu }$, and $\Phi $. Additionally, in
Section \ref{subsec:TTmassivespin2}, we argued that in order to properly
address the pathological propagation modes of the massive spin-$2$ field, it
is necessary to impose that only the transverse-traceless massive spin-$2$
modes are present in the theory. With these considerations in mind, and
utilizing $\phi = 0$, (i) equations (\ref{eq:h_plus}) and (\ref{eq:h_cross}) for the
massless spin-$2$ modes; (ii) equations (\ref{eq:Psi_plus}) and (\ref%
{eq:Psi_cross}) for the massive spin-$2$ modes, and (iii) (\ref{eq:Phi}) for
the massive spin-$0$ mode, we obtain 

\begin{itemize}
\item Massless spin-2 modes:%
\begin{equation}
h_{+}=A_{+,1}\cos\Omega t+A_{+,2}\cos2\Omega t,  \label{hplus_mode_cap3}
\end{equation}%
\begin{equation}
h_{\times}=A_{\times,1}\sin\Omega t+A_{\times,2}\sin2\Omega t,
\label{hcross_mode_cap3}
\end{equation}
where
\begin{equation}
A_{+,1}=h_{0}^{\prime}\sin2\alpha\sin\theta\cos\theta,
\end{equation}%
\begin{equation}
A_{+,2}=2h_{0}^{\prime}\sin^{2}\alpha\left( 1+\cos^{2}\theta\right) ,
\end{equation}%
\begin{equation}
A_{\times,1}=h_{0}^{\prime}\sin2\alpha\sin\theta,
\end{equation}%
\begin{equation}
A_{\times,2}=4h_{0}^{\prime}\sin^{2}\alpha\cos\theta,
\end{equation}
and%
\begin{equation}
h_{0}^{\prime}=-\frac{G}{c^{4}}\frac{\left( I_{3}-I_{1}\right) }{r}%
\Omega^{2}.  \label{h0_cap3}
\end{equation}
\begin{widetext}
\item Massive spin-2 modes:
\begin{equation}
\Psi_{+}\left( \mathbf{x},t\right) =\left\{
\begin{array}{c}
A_{+,1}\cos\left( \Omega t\right) \exp\left( -m_{\Psi}r\sqrt{1-\left( \frac{%
\Omega}{m_{\Psi}c}\right) ^{2}}\right) +A_{+,2}\cos\left( 2\Omega t\right)
\exp\left( -m_{\Psi}r\sqrt{1-\left( \frac{2\Omega}{m_{\Psi}c}\right) ^{2}}%
\right) ,\text{ \ \ }\Omega<m_{\Psi}c \\
A_{+,1}\cos\left( \Omega t\right) \exp\left( -m_{\Psi}r\sqrt{1-\left( \frac{%
\Omega}{m_{\Psi}c}\right) ^{2}}\right) +A_{+,2}\cos\left[ 2\Omega\left( t-%
\frac{r}{c}\sqrt{1-\left( \frac{m_{\Psi}c}{2\Omega}\right) ^{2}}\right) %
\right] ,\text{ \ \ }\frac{m_{\Psi}c}{2}<\Omega<m_{\Psi}c \\
A_{+,1}\cos\left[ \Omega\left( t-\frac{r}{c}\sqrt{1-\left( \frac{m_{\Psi}c}{%
\Omega}\right) ^{2}}\right) \right] +A_{+,2}\cos\left[ 2\Omega\left( t-\frac{%
r}{c}\sqrt{1-\left( \frac{m_{\Psi}c}{2\Omega}\right) ^{2}}\right) \right] ,%
\text{ \ \ }\Omega>m_{\Psi}c%
\end{array}
\right. ,  \label{psiplus_mode_cap3}
\end{equation}%
\begin{equation}
\Psi_{\times}\left( \mathbf{x},t\right) =\left\{
\begin{array}{c}
A_{\times,1}\sin\left( \Omega t\right) \exp\left( -m_{\Psi}r\sqrt{1-\left(
\frac{\Omega}{m_{\Psi}c}\right) ^{2}}\right) +A_{\times,2}\sin\left( 2\Omega
t\right) \exp\left( -m_{\Psi}r\sqrt{1-\left( \frac{2\Omega}{m_{\Psi}c}%
\right) ^{2}}\right) ,\text{ \ }\Omega<\frac{m_{\Psi}c}{2} \\
A_{\times,1}\sin\left( \Omega t\right) \exp\left( -m_{\Psi}r\sqrt{1-\left(
\frac{\Omega}{m_{\Psi}c}\right) ^{2}}\right) +A_{\times,2}\sin\left[
2\Omega\left( t-\frac{r}{c}\sqrt{1-\left( \frac{m_{\Psi}c}{2\Omega}\right)
^{2}}\right) \right] ,\text{ \ }\frac{m_{\Psi}c}{2}<\Omega<m_{\Psi}c \\
A_{\times,1}\sin\left[ \Omega\left( t-\frac{r}{c}\sqrt{1-\left( \frac{%
m_{\Psi}c}{\Omega}\right) ^{2}}\right) \right] +A_{\times,2}\sin\left[
2\Omega\left( t-\frac{r}{c}\sqrt{1-\left( \frac{m_{\Psi}c}{2\Omega}\right)
^{2}}\right) \right] ,\text{ \ \ }\Omega>m_{\Psi}c%
\end{array}
\right. ,  \label{psicross_mode_cap3}
\end{equation}
where%
\begin{align}
A_{+,1} & =\frac{1}{2}\Psi_{0}^{\prime}\sin2\alpha\sin\theta\cos\theta, \\
A_{+,2} & =\Psi_{0}^{\prime}\sin^{2}\alpha\left( 1+\cos^{2}\theta\right) .
\notag
\end{align}%
\begin{equation}
A_{\times,1}=\Psi_{0}^{\prime}\sin\alpha\cos\alpha\sin\theta,
\end{equation}%
\begin{equation}
A_{\times,2}=2\Psi_{0}^{\prime}\sin^{2}\alpha\cos\theta,
\end{equation}
and%
\begin{equation}
\Psi_{0}^{\prime}=\frac{\kappa}{4\pi r}\left( I_{3}-I_{1}\right) \Omega^{2}.
\end{equation}
\item Massive spin-0 mode:%
\begin{equation}
\Phi^{Q}\left( \mathbf{x},t\right) =\left\{
\begin{array}{c}
A_{s,1}\cos\left( \Omega t\right) \exp\left( -m_{\Phi}r\sqrt{1-\left( \frac{%
\Omega}{m_{\Phi}c}\right) ^{2}}\right) +A_{s,2}\cos\left( 2\Omega t\right)
\exp\left( -m_{\Phi}r\sqrt{1-\left( \frac{2\Omega}{m_{\Phi}c}\right) ^{2}}%
\right) ,\text{ \ \ }\Omega<m_{\Phi}c \\
A_{s,1}\cos\left( \Omega t\right) \exp\left( -m_{\Phi}r\sqrt{1-\left( \frac{%
\Omega}{m_{\Phi}c}\right) ^{2}}\right) +A_{s,2}\cos\left[ 2\Omega\left( t-%
\frac{r}{c}\sqrt{1-\left( \frac{m_{\Phi}c}{2\Omega}\right) ^{2}}\right) %
\right] ,\text{ \ \ }\frac{m_{\Phi}c}{2}<\Omega<m_{\Phi}c \\
A_{s,1}\cos\left[ \Omega\left( t-\frac{r}{c}\sqrt{1-\left( \frac{m_{\Phi}c}{%
\Omega}\right) ^{2}}\right) \right] +A_{s,2}\cos\left[ 2\Omega\left( t-\frac{%
r}{c}\sqrt{1-\left( \frac{m_{\Phi}c}{2\Omega}\right) ^{2}}\right) \right] ,%
\text{ \ \ }\Omega>m_{\Phi}c%
\end{array}
\right. ,  \label{Psi_Q_final}
\end{equation}
where%
\begin{equation}
A_{s,1}=\Phi_{0}^{\prime}\sin\theta\cos\theta\sin2\alpha,  \notag
\end{equation}%
\begin{equation*}
A_{s,2}=\Phi_{0}^{\prime}\sin^{2}\theta\left( \cos2\alpha-1\right) ,
\end{equation*}
\end{widetext}
and%
\begin{equation*}
\Phi_{0}^{\prime}=\frac{1}{3}\frac{G}{c^{4}}\frac{\left( I_{3}-I_{1}\right)
}{r}\Omega^{2}.
\end{equation*}
\end{itemize}

The resulting gravitational waves for all fields are a combination of two
frequency spectra: $\Omega $ and $2\Omega $. The presence of these two
spectra provides the following interpretation for the damping and
oscillatory modes of the massive fields: Initially, when $\Omega <m_{X}c$,
both waves (with frequencies $\omega _{gw}=\Omega $ and $\omega
_{gw}=2\Omega $) are in damping modes. As $\Omega $ increases to $\Omega >%
\frac{m_{X}c}{2}$, the wave with frequency $\omega _{gw}=2\Omega $
transitions to the oscillatory mode, while the wave with $\omega
_{gw}=\Omega $ remains in the damping mode. Finally,when $\Omega >m_{X}c$,
the wave with $\omega _{gw}=\Omega $ also enters the oscillatory mode,
completing the waveform for the corresponding massive mode. Note that in the
full oscillatory regime, the solutions $h_{+,\times }$ and $\Psi _{+,\times
} $ are waves with the same amplitude and frequency but with different wave
numbers. This allows for an interpretation of destructive interference
effects, as discussed in \cite{Alves2022}. Furthermore, in the limit of $%
m_{\Psi }\rightarrow 0$, this destructive interference becomes complete,
resulting in the absence of tensorial mode emission.

Using the equations above, we can calculate the loss of energy and angular
momentum through Eqs. (\ref{eq:energy-loss-TT}) and (\ref%
{eq:angular-momentum-loss-TT}). By substituting the propagation modes into (%
\ref{eq:energy-loss-TT}), we obtain the energy loss as

\begin{align}
\frac{dE}{dt}& =A_{E}\left\{ 16\sin ^{2}\alpha \left[ 1-\sqrt{1-\left( \frac{%
m_{\Psi }c}{2\Omega }\right) ^{2}}\right. \right.   \notag
\\
& \left. +\frac{1}{18}\sqrt{1-\left( \frac{m_{\Phi }c}{2\Omega }\right) ^{2}}%
\right] +\cos ^{2}\alpha \left[ 1-\sqrt{1-\left( \frac{m_{\Psi }c}{2\Omega }%
\right) ^{2}}\right.   \notag \\
& \left. \left. +\frac{1}{18}\sqrt{1-\left( \frac{m_{\Phi }c}{\Omega }%
\right) ^{2}}\right] \right\} , \label{energy_balance_eq}
\end{align}%
with%
\begin{equation}
A_E=-\frac{2G}{5c^{5}}\left( I_{1}-I_{3}\right) ^{2}\Omega ^{6}\sin ^{2}\alpha 
\end{equation}%
and by substituting them into (\ref{eq:angular-momentum-loss-TT}), we get
the angular momentum loss:%
\begin{align}
\frac{dJ}{dt}& =A_J\left\{ 16\sin ^{2}\alpha \left[ 1-\sqrt{1-\left( \frac{%
m_{\Psi }c}{2\Omega }\right) ^{2}}\right. \right.  \notag
\\
& \left. +\frac{1}{18}\sqrt{1-\left( \frac{m_{\Phi }c}{2\Omega }\right) ^{2}}%
\right] +\cos ^{2}\alpha \left[ 1-\sqrt{1-\left( \frac{m_{\Psi }c}{\Omega }%
\right) ^{2}}\right.   \notag \\
& \left. \left. +\frac{1}{18}\sqrt{1-\left( \frac{m_{\Phi }c}{\Omega }%
\right) ^{2}}\right] \right\} ,  \label{angular_momentum}
\end{align}%
with%
\begin{equation}
A_{J}=-\frac{2G}{5c^{5}}\left( I_{1}-I_{3}\right) ^{2}\Omega ^{5}\sin
^{2}\alpha 
\end{equation}%
By comparing this result with Eq. (\ref{energy_balance_eq}), we find that $%
\frac{dE}{dt}=\Omega \frac{dJ}{dt}$. Since $\Omega =\dot{\beta}$ and $J=I_{1}%
\dot{\beta}$ by Eq. (\ref{omega_beta}), the above equation gives
\begin{align}
\ddot{\beta}& =A_{\beta }\left\{ 16\sin ^{2}\alpha \left[ 1-\sqrt{1-\left( 
\frac{m_{\Psi }c}{2\Omega }\right) ^{2}}\right. \right. \notag \\
& \left. +\frac{1}{18}\sqrt{1-\left( \frac{m_{\Phi }c}{2\Omega }\right) ^{2}}%
\right] +\cos ^{2}\alpha \left[ 1-\sqrt{1-\left( \frac{m_{\Psi }c}{\Omega }%
\right) ^{2}}\right.   \notag \\
& \left. \left. +\frac{1}{18}\sqrt{1-\left( \frac{m_{\Phi }c}{\Omega }%
\right) ^{2}}\right] \right\} , \label{beta_equation}
\end{align}%
with%
\begin{equation}
A_{\beta }=-\frac{2G}{5c^{5}}\frac{\left( I_{1}-I_{3}\right) ^{2}}{I_{1}}%
\Omega ^{5}\sin ^{2}\alpha .
\end{equation}

A similar analysis can be conducted to determine the evolution of the angle $%
\alpha$. In this case, the equation governing the evolution of $\alpha$ is
given by:
\begin{align}
\dot{\alpha}& =A_{\alpha }\left\{ 16\sin ^{2}\alpha \left[ 1-\sqrt{1-\left( 
\frac{m_{\Psi }c}{2\dot{\beta}}\right) ^{2}}\right. \right. 
\notag  \\
& \left. +\frac{1}{18}\sqrt{1-\left( \frac{m_{\Phi }c}{2\dot{\beta}}\right)
^{2}}\right] +\cos ^{2}\alpha \left[ 1-\sqrt{1-\left( \frac{m_{\Psi }c}{\dot{%
\beta}}\right) ^{2}}\right.   \notag \\
& \left. \left. +\frac{1}{18}\sqrt{1-\left( \frac{m_{\Phi }c}{\dot{\beta}}%
\right) ^{2}}\right] \right\} , \label{alpha_equation}
\end{align}%
with%
\begin{equation}
A_{\alpha }=-\frac{2G}{5c^{5}}\frac{\left( I_{1}-I_{3}\right) ^{2}}{I_{1}}%
\dot{\beta}^{4}\sin \alpha \cos \alpha .
\end{equation}%
In order to examine the coupled system given by Eqs. (\ref{beta_equation})
and (\ref{alpha_equation}) in greater detail, we introduce the dimensionless
function as given by
\begin{equation}
\omega\left( t\right) =\frac{\dot{\beta}\left( t\right) }{\dot{\beta}_{0}},
\end{equation}
where $\beta_{0}$ represents the value of $\beta\left( t=0\right) $.
Additionally, we introduce the corresponding time scale

\begin{equation}
\tau_{0}=\left[ \frac{2G}{5c^{5}}\frac{\left( I_{1}-I_{3}\right) ^{2}}{I_{1}}%
\dot{\beta}_{0}^{4}\right] ^{-1}\text{ }.
\end{equation}
Using these new definitions, we can rewrite equations (\ref{beta_equation})
and (\ref{alpha_equation}) as follows:
\begin{align}
\dot{\omega}& =-\frac{1}{\tau _{0}}\omega ^{5}\sin ^{2}\alpha \left\{ 16\sin
^{2}\alpha \left[ 1-\sqrt{1-\left( \frac{m_{\Psi }c}{2\omega \dot{\beta}_{0}}%
\right) ^{2}}\right. \right.   \notag  \\
& \left. +\frac{1}{18}\sqrt{1-\left( \frac{m_{\Phi }c}{2\omega \dot{\beta}%
_{0}}\right) ^{2}}\right] +\cos ^{2}\alpha \left[ 1-\sqrt{1-\left( \frac{%
m_{\Psi }c}{\omega \dot{\beta}_{0}}\right) ^{2}}\right.   \notag \\
& \left. \left. +\frac{1}{18}\sqrt{1-\left( \frac{m_{\Phi }c}{\omega \dot{%
\beta}_{0}}\right) ^{2}}\right] \right\},  \label{uEq}
\end{align}%
and%
\begin{align}
\dot{\alpha}& =-\frac{1}{\tau _{0}}\omega ^{4}\sin \alpha \cos \alpha
\left\{ 16\sin ^{2}\alpha \left[ 1-\sqrt{1-\left( \frac{m_{\Psi }c}{2\omega 
\dot{\beta}_{0}}\right) ^{2}}\right. \right.   \notag \\
& \left. +\frac{1}{18}\sqrt{1-\left( \frac{m_{\Phi }c}{2\omega \dot{\beta}%
_{0}}\right) ^{2}}\right] +\cos ^{2}\alpha \left[ 1-\sqrt{1-\left( \frac{%
m_{\Psi }c}{\omega \dot{\beta}_{0}}\right) ^{2}}\right.   \notag \\
& \left. \left. +\frac{1}{18}\sqrt{1-\left( \frac{m_{\Phi }c}{\omega \dot{%
\beta}_{0}}\right) ^{2}}\right] \right\} .  \label{AlphaEq}
\end{align}

\begin{figure}[h!]
\begin{centering}
\includegraphics[scale=0.7]{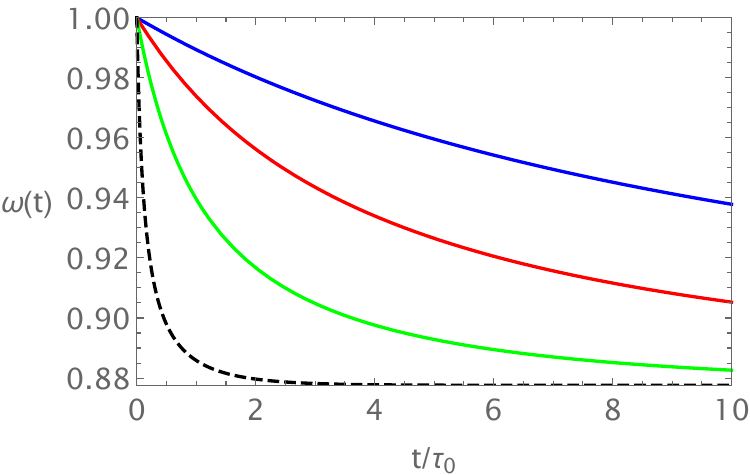}
\par\end{centering}
\caption{Dimensionless precession velocity as a function of time (in $%
\protect\tau_{0}$ units) for different values of $m_{\Psi }c$. The blue, red and green curves
are constructed with $m_{\Psi}c=0.3$, $m_{\Psi}c=0.5$ and $m_{\Psi}c=0.9$,
respectively. The dashed black curve represents the GR solution.}
\label{fig1}
\end{figure}

\begin{figure}[h!]
\begin{centering}
\includegraphics[scale=0.7]{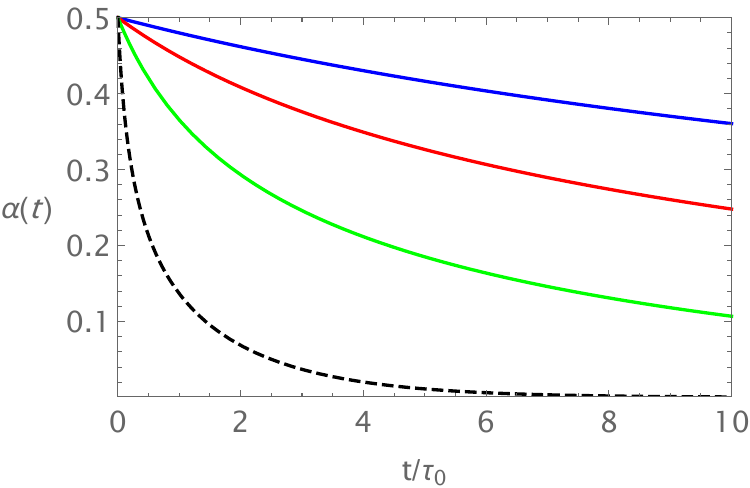}
\par\end{centering}
\caption{Wobble angle as a function of time (in $\protect\tau_{0}$ units)
for different values of $m_{\Psi }c$. The blue, red and green curves
are constructed with $m_{\Psi}c=0.3$, $m_{\Psi}c=0.5$ and $m_{\Psi}c=0.9$,
respectively. The dashed black curve represents the GR solution.}
\label{fig2}
\end{figure}

To analyze the impact of massive modes on rotational dynamics, we
will focus our study specifically on the tensor mode within the oscillatory
regime. In this case, by integrating Eqs. (\ref{uEq}) and (\ref{AlphaEq}), we
get the curves that are shown in Figs. \ref{fig1} and \ref{fig2}. Figure \ref%
{fig1} clearly shows that the frequency decreases with time, similar to what is predicted by GR. However, it also shows that the inclusion of a
massive field moderates this frequency reduction, causing it to decrease
more slowly as the mass $m_{\Psi }$ decreases. This happens because the
reduction in $m_{\Psi }$ increases the destructive interference between the
fields $h_{\mu \nu }$ and $\Psi _{\mu \nu }$, leading to a slower loss of
energy by the system. A similar effect was observed in \cite{Alves2022} for
point-like binary mass systems. The greater effectiveness of destructive
interference with decreasing mass can be understood by recalling that $%
m_{\Psi }\propto \frac{1}{\alpha }$, where $\alpha $ is the coupling
constant associated with the $C^{2}$ term in the action. Therefore, as $%
m_{\Psi }$ decreases, $\alpha $ increases, amplifying the impact of the
massive spin-$2$ field.

Similarly, Figure \ref{fig2} shows that the angular inclination $\alpha $
decreases over time, such as GR predictions. Once again, the presence of
the massive field slows this decrease, and the reduction in angular
inclination becomes progressively slower as the destructive interference
becomes more effective.

\section{Final comments\label{sec:Final-comments}}

In this paper, we analyzed the emission of gravitational waves within the framework of quadratic gravity, including the contribution of longitudinal and transverse massive modes, extending the previous analysis presented in Ref. \cite{Alves2022}. Firstly, we derived the wave equations for the massless ($\tilde{h}_{\alpha\beta}$) and massive ($\Phi, \Psi_{\alpha\beta}$) perturbations, followed by obtaining the associated conserved currents. Subsequently, we calculated the energy and angular momentum radiated through gravitational waves. Focusing on the quadrupole approximation, we examined two particular cases: a binary point-mass system in circular orbit, and a rotating ellipsoid with free precession.

An important point we demonstrate in this work is that, under the quadrupole approximation, only the longitudinal modes associated with the massive spin-2 term exhibit unavoidable issues, such as negative power emission. When we restrict our analysis to only the transverse-traceless modes, the destructive interference between massless and massive spin-2 perturbations yields a consistent model that ensures the positivity of both the energy and angular momentum carried by the gravitational waves. 

Considering only the contribution of the transverse-traceless modes of the massive tensorial perturbation, we analyze the interference effects between the massive and massless spin-2 modes for the rotating ellipsoid case. Our results indicate that the presence of $\Psi_{\alpha\beta}$ softens the decrease in both the angular velocity and wobble angle of the ellipsoid.

The presence of modes with negative energy is a typical feature of higher-order theories, and in the present case, they arise from the term quadratic in the Weyl tensor in the Lagrangian. The restriction proposed here is quite appealing in the sense that it allows us to preserve the term $C^{\mu\nu\alpha\beta}C_{\mu\nu\alpha\beta}$ while still ensuring the positivity of the energy. This approach may eventually be employed in other contexts, e.g. quantum gravity, and hopefully it may contribute to eliminating, or at least controlling, the ghost instabilities that usually plague higher-order theories. This is a subject that shall be explored elsewhere in a future work.

\section*{Acknowledgements}

MFSA thanks FAPES (Brazil) for its support. The authors RRC and LGM thank CNPq-Brazil for partial financial support—Grants: 309063/2023-0 (RRC) and 307901/2022-0 (LGM). RRC and CAMM are grateful to FAPEMIG-Brazil (Grants APQ-00544- 23 and APQ-05218-23) for partial financial support. PJP acknowledges LAB-CCAM at ITA and CNPq (Grant 401565/2023-8). The authors are grateful to the referee whose comments helped to improve the paper.

\appendix

\section{Conserved current\label{app:Current}}

The conserved current associated with a general action $S$ can be obtained
by calculating the total variation $\delta$ with respect to this action.

The total variation $\delta$ is both spacetime dependent, 
\begin{equation}
x^{\prime\mu}=x^{\mu}+\delta x^{\mu},  \label{eq:x_prime}
\end{equation}
and field dependent, 
\begin{equation}
\phi_{i}^{\prime}\left(x^{\prime}\right)=\phi_{i}\left(x\right)+\delta%
\phi_{i}\left(x\right).  \label{eq:phi_prime}
\end{equation}
By defining a new variation operator $\bar{\delta}$ which acts only in the
functional form of $\phi_{i}\left(x\right)$, 
\begin{equation}
\bar{\delta}\phi_{i}\left(x\right)\equiv\phi_{i}^{\prime}\left(x\right)-%
\phi_{i}\left(x\right),  \label{eq:delta_bar_phi}
\end{equation}
it is possible to split the total variation as 
\begin{equation}
\delta\phi_{i}\left(x\right)=\bar{\delta}\phi_{i}\left(x\right)+\delta
x^{\mu}\partial_{\mu}\phi_{i}\left(x\right).  \label{eq:delta_phi}
\end{equation}
Moreover, by construction 
\begin{equation}
\left[\partial_{\mu},\bar{\delta}\right]\phi_{i}\left(x\right)=0.
\label{eq:commut(del,delta_bar)}
\end{equation}

Let us start with a second-order general action given by 
\begin{equation}
S=\int d^{4}x\mathcal{L}\left[\phi_{i}\left(x\right),\partial_{\mu}\phi_{i}%
\left(x\right),\partial_{\nu}\partial_{\mu}\phi_{i}\left(x\right)\right],
\label{eq:A}
\end{equation}
where $\phi_{i}$ represents a collective set of fields with $i=1,2,\dots,N$.

The first step is to take the total variation of the action $S$ in (\ref%
{eq:A}): 
\begin{align}
\delta S & = \int\left[\delta\left(d^{4}x\right)\mathcal{L}+d^{4}x\delta\mathcal{
L}\right] \notag \\
& =\int d^{4}x\left[{\partial_{\mu}\left(\delta x^{\mu}\right)%
\mathcal{L}}+\delta\mathcal{L}\right].
\label{eq:delta_A(d(delta_x),delta_L)}
\end{align}
The (total) variation of the Lagrangian, $\delta\mathcal{L}$, will have a
contribution from the coordinate variation and another from the variations
in the functional form of the fields: 
\begin{equation}
\delta\mathcal{L}=\bar{\delta}\mathcal{L}+\delta x^{\mu}\partial_{\mu}%
\mathcal{L}.  \label{eq:delta_L}
\end{equation}

Now, 
\begin{align}
\bar{\delta}\mathcal{L} &=\frac{\partial \mathcal{L}}{\partial \phi _{i}}%
\bar{\delta}\phi _{i}+\frac{\partial \mathcal{L}}{\partial \left( \partial
_{\mu }\phi _{i}\right) }\bar{\delta}\left( \partial _{\mu }\phi _{i}\right) \notag \\
&+\frac{\partial \mathcal{L}}{\partial \left( \partial _{\nu }\partial
_{\mu }\phi _{i}\right) }\bar{\delta}\left( \partial _{\nu }\partial _{\mu
}\phi _{i}\right) . \label{eq:delta_bar_L}
\end{align}%
Eq. (\ref{eq:commut(del,delta_bar)}) allow us to commute $\bar{\delta}$ and $%
\partial _{\mu }$. Thus, commuting these operators and applying the Leibniz
rule a few times, we obtain 
\begin{align}
\bar{\delta}\mathcal{L} &=\frac{\delta \mathcal{L}}{\delta \phi _{i}}\bar{\delta}\phi _{i}+\partial _{\mu }\partial _{\nu }{\left[ \frac{\partial 
\mathcal{L}}{\partial \left( \partial _{\mu }\partial _{\nu }\phi
_{i}\right) }\bar{\delta}\phi _{i}\right] }  \notag 
\\
&+\partial _{\mu }{\left[ \frac{\partial \mathcal{L}}{\partial \left(
\partial _{\mu }\phi _{i}\right) }-2\partial _{\nu }\frac{\partial \mathcal{L%
}}{\partial \left( \partial _{\mu }\partial _{\nu }\phi _{i}\right) }\right] 
}\bar{\delta}\phi _{i}, \label{eq:delta_bar_L_Noether}
\end{align}%
where 
\begin{equation}
{\frac{\delta \mathcal{L}}{\delta \phi _{i}}\equiv \frac{\partial \mathcal{L}%
}{\partial \phi _{i}}-\partial _{\mu }\frac{\partial \mathcal{L}}{\partial
\left( \partial _{\mu }\phi _{i}\right) }+\partial _{\mu }\partial _{\nu }%
\frac{\partial \mathcal{L}}{\partial \left( \partial _{\nu }\partial _{\mu
}\phi _{i}\right) }.}  \label{eq:delta_L_delta_phi}
\end{equation}%
By substituting Eqs. (\ref{eq:delta_bar_L_Noether}) and (\ref{eq:delta_L})
in Eq. (\ref{eq:delta_A(d(delta_x),delta_L)}), we get 
\begin{align}
\delta S &= \int d^{4}x\left\{ {\frac{\delta \mathcal{L}}{\delta \phi _{i}}}
\bar{\delta}\phi _{i}+\partial _{\mu }{\left[ \frac{\partial \mathcal{L}}{
\partial \left( \partial _{\mu }\phi _{i}\right) }-2\partial _{\nu }\frac{
\partial \mathcal{L}}{\partial \left( \partial _{\mu }\partial _{\nu }\phi
_{i}\right) }\right] }\bar{\delta}\phi _{i}\right. 
\notag \\
& \left. +\partial _{\mu }\left( {\mathcal{L}\delta x^{\mu }}\right)
+\partial _{\mu }\partial _{\nu }\left( {\frac{\partial \mathcal{L}}{
\partial \left( \partial _{\mu }\partial _{\nu }\phi _{i}\right) }}\bar{
\delta}\phi _{i}\right) \right\}.  \label{eq:delta_A_Noether_v2}
\end{align}

Suppose that the total variation is generated by a transformation
characterized by a set of parameters $\epsilon ^{a}$. In this case,
the equations (\ref{eq:x_prime}) and (\ref{eq:phi_prime}) assume the form 
\begin{align*}
& \delta x^{\mu } =\epsilon ^{a}A_{a}^{\mu }\left( x\right) ,\text{ \ }\delta
\phi _{i}\left( x\right) =\epsilon ^{a}F_{i,a}\left( \phi ,\partial \phi
,\partial ^{2}\phi \right) \Rightarrow \\
& \Rightarrow \bar{\delta}\phi _{i}\left( x\right) =\epsilon ^{a}\left[ F_{i,a}\left(
\phi ,\partial \phi ,\partial ^{2}\phi \right) -A_{a}^{\sigma }\left(
x\right) \partial _{\sigma }\phi _{i}\right] .
\end{align*}%
In this notation, Eq. (\ref{eq:delta_A_Noether_v2}) can be written as\ 
\begin{equation}
\delta S=\int d^{4}x\left[ {\frac{\delta \mathcal{L}}{\delta \phi _{i}}}%
\epsilon ^{a}\left[ F_{i,a}-A_{a}^{\sigma }\partial _{\sigma }\phi _{i}%
\right] -\epsilon ^{a}\partial _{\mu }j_{a}^{\mu }\right] ,
\label{eq:delta_A_Noether(j)}
\end{equation}%
where 
\begin{eqnarray}
j_{a}^{\mu } &=&{\left[ \frac{\partial \mathcal{L}}{\partial \left( \partial
_{\mu }\phi _{i}\right) }-2\partial _{\nu }\frac{\partial \mathcal{L}}{%
\partial \left( \partial _{\mu }\partial _{\nu }\phi _{i}\right) }\right] }%
\left[ A_{a}^{\sigma }\partial _{\sigma }\phi _{i}-F_{i,a}\right] 
\notag \\
&&-{A_{a}^{\mu }\left( x\right) \mathcal{L+}}\partial _{\nu }\left\{ {\frac{%
\partial \mathcal{L}}{\partial \left( \partial _{\mu }\partial _{\nu }\phi
_{i}\right) }}\left[ A_{a}^{\sigma }\partial _{\sigma }\phi _{i}-F_{i,a}%
\right] \right\}  \notag
\end{eqnarray}
is the definition of Noether's current for our higher-order Lagrangian.

Note that when the transformation considered is a symmetry of the action ($%
\delta S=0$), the solutions of the field equations $\phi_{i}^{\rm{sol}}$ which
satisfy the Euler-Lagrange equation 
\begin{equation*}
\frac{\delta\mathcal{L}}{\delta\phi_{i}^{\rm{sol}}}=0,
\end{equation*}
lead to the conservation of the current, as follows: 
\begin{equation}
\left.\partial_{\mu}j_{a}^{\mu}\right\vert _{\phi_{i}=\phi_{i}^{\rm{sol}}}=0.
\label{eq:CC}
\end{equation}

\bibliographystyle{apsrev4-1}
\bibliography{main}

\end{document}